\documentclass[journal=jctcce,manuscript=article,email=True]{achemso}

\usepackage[T1]{fontenc} 
\usepackage[alsoload=synchem]{siunitx}
\usepackage{amsmath}

 \DeclareSIUnit\pn{\pico\newton}
 \DeclareSIUnit[per-mode=symbol]\pns{\pico\newton\per\second}
 \DeclareSIUnit[per-mode=symbol]\nm{\nano\metre}
 \DeclareSIUnit[per-mode=symbol]\nms{\nano\metre\per\second}
 \DeclareSIUnit[per-mode=symbol]\pnnm{\pico\newton\per\nano\metre}
 \DeclareSIUnit{\microsecond}{\SIUnitSymbolMicro s}
 \DeclareSIUnit\M{\Molar}

\author{Megan C. Engel}
\affiliation[Harvard]
{School of Engineering and Applied Sciences, Harvard University, 29 Oxford Street, Cambridge MA, 02138, USA}
\alsoaffiliation{Rudolf Peierls Centre for Theoretical Physics, University of Oxford, 1 Keble Road, Oxford, OX1 3NP, UK}
\email{mcengel@seas.harvard.edu}
\author{Flavio Romano}
\affiliation[Venice]
{Dipartimento di Scienze Molecolari e Nanosistemi, Universitá Ca Foscari di Venezia, Via Torino 155, 30172, Venezia Mestre, Italy}
\author{Ard. A. Louis}
\affiliation[OxPhys]
{Rudolf Peierls Centre for Theoretical Physics, University of Oxford, 1 Keble Road, Oxford, OX1 3NP, UK}
\author{Jonathan P. K. Doye}
\affiliation[OxChem]
{Department of Chemistry, University of Oxford, South Parks Road, Oxford, OX1 3QZ, UK}

\title[]
  {Measuring internal forces in single-stranded DNA: Application to a DNA force clamp}

\abbreviations{IR,NMR,UV}

\begin{document}







\begin{abstract}
We present a new method for calculating internal forces in DNA structures using coarse-grained models and demonstrate its utility with the oxDNA model. The instantaneous forces on individual nucleotides are explored and related to model potentials, and using our framework, internal forces are calculated for two simple DNA systems and for a recently-published nanoscopic force clamp. Our results highlight some pitfalls associated with conventional methods for estimating internal forces, which are based on elastic polymer models, and emphasise the importance of carefully considering secondary structure and ionic conditions when modelling the elastic behaviour of single-stranded DNA. Beyond its relevance to the DNA nanotechnological community, we expect our approach to be broadly applicable to calculations of internal force in a variety of structures -- from DNA to protein -- and across other coarse-grained simulation models.
\end{abstract}

\section{Introduction}


The communication of force is central to cellular function. Force propagation through DNA features in transcription and translation, as special molecules forcibly open the DNA helix to access its contents~\cite{book:Alberts2008,Kramm2020}; in the dense packaging that enables $\sim$metres of DNA to be stored in a $\sim$6 \si{\micro\metre}-diameter nucleus\cite{book:Alberts2008}; and even in gene expression, which is is modulated by the mechanical state of the nucleus~\cite{Uhler2017}. Force plays a role in manifold other biological mechanisms, as well: mechanosensitive ion channels open in response to specific tension in cellular membranes~\cite{Ingber2006,Arnadottir2010};  the catalytic activity of enzymes can be altered by the application of force~\cite{Ingber2006}; and the cytoskeleton and extracellular matrix are tensegrity networks that propagate forces within, between, and among cells~\cite{Wang2017}. Additionally, force and torque are important to the functioning of molecular machines, from the transport molecule kinesin to ATP synthase, the energy production factory of all cells~\cite{Bustamante2004}. Because of their critical role in diverse processes, from embryogenesis~\cite{Poh2014} to immune response~\cite{Kim2009,Spillane2017} and tumour metastasis~\cite{Levental2009}, probing forces at the molecular level is a key scientific aim.

The advent of DNA nanotechnology~\cite{Seeman1982,Seeman2018} has enabled the fabrication of DNA-based nanoscale devices capable of sensing~\cite{GuSeeman2010,Cost2015,Blanchard2019} and exerting~\cite{Nickels2016,Thubagere2017,Blanchard2019,Kramm2020} forces in the biologically relevant regime of 0-20\,\si{\pn}~\cite{Shroff2005}. DNA force sensors have already been profitably applied to study integrin proteins that connect cells to extracellular filaments~\cite{Brockman2018}; the forces involved in nucleosome association\cite{Funke2016}; and DNA repair proteins~\cite{GuSeeman2010}. Force-generating DNA devices have been used to investigate the binding of transcription factors under tension~\cite{Nickels2016,Kramm2020} and to mimic the function of molecular motors in transporting cargo~\cite{Thubagere2017}. These devices exploit the mechano-elastic properties of DNA in its single- and double-stranded forms. Some sense forces using the known rupture forces of double-stranded DNA (dsDNA) in unzipping or shearing geometries~\cite{Ha2013}. Others use the unfolding thermodynamics of DNA hairpins~\cite{Blakely2014,Zhang2014,Dutta2018,Liu2016,Ma2019}: when hairpins unfold in response to an applied load, it is inferred that the force exceeds F$_{1/2}$, the force at which a hairpin has a 50\% chance of being unfolded at equilibrium. In computing F$_{1/2}$, elastic polymer models are typically employed to calculate the free energy of stretching single-stranded DNA (ssDNA). Polymer models are also invoked to calibrate nanodevices that capitalize on the \emph{entropic spring} behaviour of ssDNA, which resists any decrease in configurational entropy: circumscribing the end-to-end distance of ssDNA by even a few \si{\nm} leads to forces in the \si{\pn} range~\cite{Smith1995,Camunas-Soler2016}. Sensors use the end-to-end distances of ssDNA to report forces~\cite{Shroff2005,Funke2016}, and other devices exert predefined forces on targets by constraining the end-to-end distance of ssDNA~\cite{Nickels2016,Zocchi2009,Meng2011,Kramm2020}. 

Many nanotechnological applications beyond force sensors and probes require knowledge of the mechano-elastic properties of ssDNA: in so-called DNA wireframe architectures like polyhedra and buckyballs~\cite{Goodman2005,HeMao2008}, flexible ssDNA strands act as vertices that link stiff double-stranded regions to achieve designed shapes. Some tensegrity structures -- like DNA origami~\cite{Rothemund2006} tensegrity prisms and kites -- feature ssDNA as the mediator of constant stress in the design~\cite{Liedl2010,Simmel2014}.
\citeauthor{Liedl2010} demonstrated the ability of DNA helix bundles in these structures to assemble against a force of $\sim$14\,\si{\pn} applied by ssDNA linkers. 
ssDNA springs have been used to introduce flexibility into DNA origami crank-slider joints~\cite{Marras2015}; to act as hinges in nanocages designed for drug delivery~\cite{Douglas2012}; and as components of nanotubes capable of the controlled release of gold nanoparticles~\cite{Lo2010}. Finally, many computational models utilised to refine and design DNA nanostructures employ polymer models of ssDNA~\cite{Kim2012,Reshetnikov2018}, rendering the importance of accurate mechanical models of DNA far-reaching.

To determine the forces exerted on or applied by ssDNA in biological and nanotechnological contexts, two general routes have been employed: experimental calibration and simplified polymer models~\cite{Cost2015}. In the former case, single-molecule force spectroscopy (SMFS), for example via magnetic tweezers, is used to characterise the force response of the tension sensor in question~\cite{Shroff2005,Dutta2018}. SMFS experiments are prone to instrumental artefacts, however, including convolution of tension probe with the instrument response function and the fact that bond rupture forces depend on loading rates\cite{Evans1997,Bullerjahn2014} used during calibration. These issues complicate the interpretation of forces and lead to uncertainties of at least $\sim$a few\,\si{\pn}~\cite{Shroff2005,Dutta2018}. 


\begin{figure}
\begin{center}
\includegraphics[width=\textwidth]{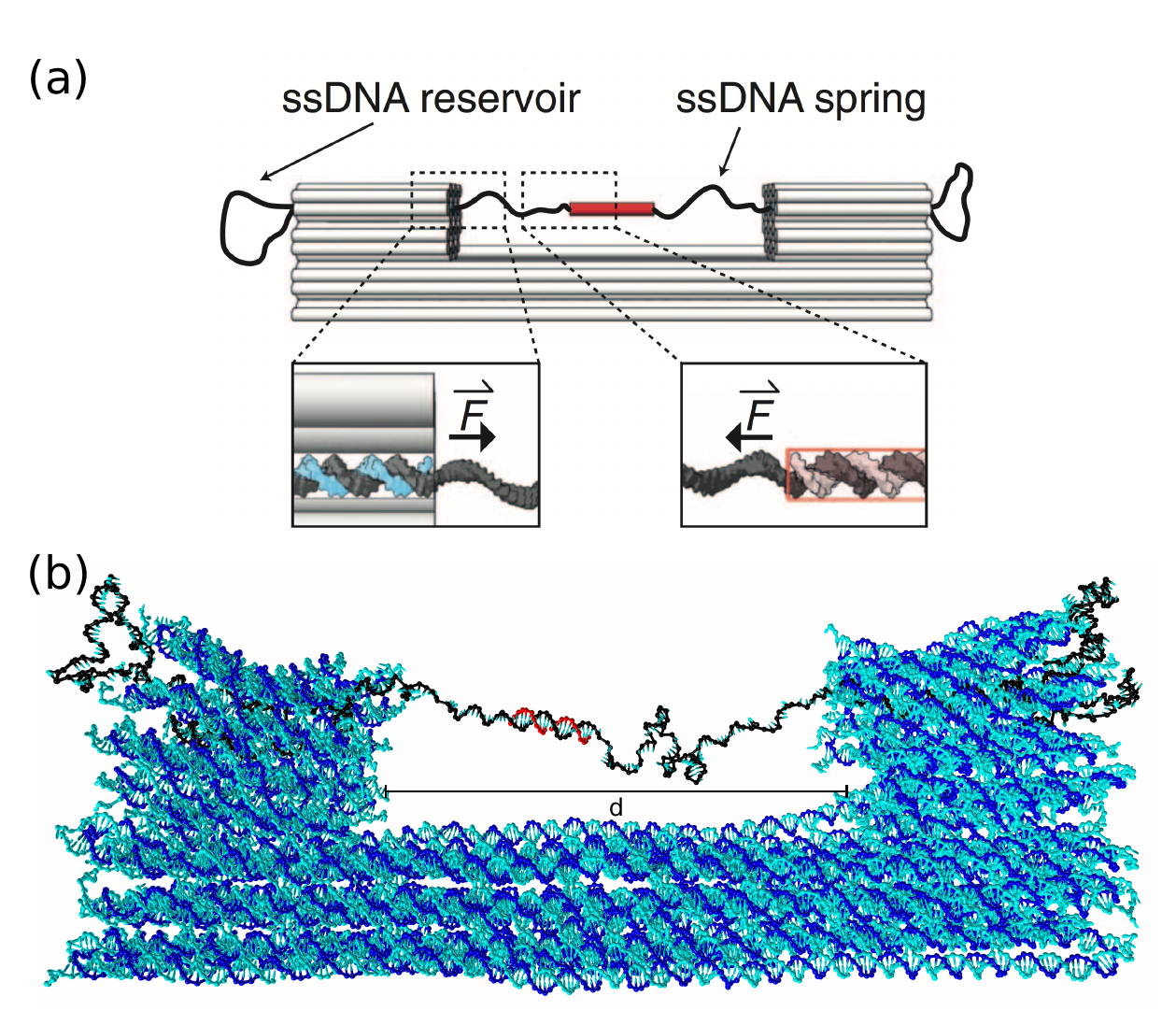}
\caption[]{Illustration of the force clamp design of \citet{Nickels2016}. (a) Schematic and (b) oxDNA representation of the force clamp: a portion of ssDNA to which a system of interest is bound (red) is suspended in a DNA origami frame. Its length can be adjusted by altering staple sequences such that material in the ssDNA `reservoirs' (black) shifts in or out of the central gap. (a) is adapted with permission from \citet{Nickels2016}. The designed gap distance $d$ is 42.84\,\si{\nm}.}
\label{fig:fullnickels}
\end{center}
\end{figure}
Alternatively, because ssDNA has been widely described by simplified polymer models -- typically the worm-like chain (WLC)~\cite{Odijk1995} or the extensible freely-jointed chain (exFJC)~\cite{book:Rubinstein2003} model -- the majority of DNA tension probes are calibrated by applying such models with previously-measured parameters. 
%

One example of a DNA tension probe calibrated using simplified polymer models is the recently-proposed DNA origami force clamp of \citet{Nickels2016}, shown in Figure~\ref{fig:fullnickels}. Force is applied to a molecule of interest bound to the centre of a ssDNA spring suspended between rigid origami blocks; the magnitude of this force can be tuned by adjusting the spring's length. In practice, this is achieved by altering the internal origami staple sequences slightly to thread portions of ssDNA `reservoirs' on either end of the origami into or out of the central gap. As we will detail below, potential problems with common calibration procedures like the one employed by \citet{Nickels2016} include the fact that it is not clear how to map ssDNA straightforwardly to existing polymer models, as well as the existence of secondary structures in ssDNA which they do not account for. 

Given the challenge of calculating ssDNA internal forces precisely and the relevance of doing so to manifold biological and nanotechnological applications, simulations have a clear role in complementing and refining experiments. However, an explicit procedure for calculating these forces in coarse-grained simulations has not been delineated elsewhere. Here, we present a novel method for calculating internal forces in coarse-grained simulations. We then use this framework to analyse the nanoscopic force clamp of \citet{Nickels2016}, which allows us to highlight crucial pitfalls associated with using continuum polymer models to calibrate internal forces in ssDNA.

To achieve our end of providing accurate internal force estimates in a DNA nanostructure, a computational model of DNA is needed that is detailed enough to capture DNA mechanics well, but sufficiently coarse-grained that it can be simulated on long enough time scales to (i) sample all relevant ssDNA secondary structure and (ii) achieve a converged average over instantaneous forces that may have large variance. Here, we use oxDNA~\cite{Ouldridge2011,Sulc2012,Snodin2015}, a single-nucleotide-level coarse grained DNA model that has enjoyed extensive validation at multiple levels of description. Basic DNA biophysical properties, including structure, persistence length~\cite{Ouldridge2011,Snodin2015,Sulc2012}, torsional modulus~\cite{Matek2015,Skoruppa2018}, and the force-extension curves of single- and double-stranded DNA~\cite{Sulc2012} are well-captured by oxDNA, which also successfully modelled force-induced dsDNA overstretching~\cite{RomanoOverstretching2013}. Additionally, oxDNA has successfully described large DNA origami designs both structurally -- capturing the opening angles of DNA origami hinges~\cite{Sharma2017,Shi2017} and the full 3D structure of a cryo-EM characterised origami~\cite{Snodin2019} -- and mechanically, reproducing experimental origami force unfolding behaviour~\cite{Engel2018}. Finally, oxDNA has repeatedly been shown to capture the thermodynamics of DNA hybridization and hairpin formation reactions well~\cite{Ouldridge2011,Sulc2012,Snodin2015}. Since these are the crucial processes relevant to ssDNA secondary structure formation, we are confident that oxDNA is capable of sampling secondary structures realistically; indeed, previous studies have validated this ability explicitly through deriving ssDNA secondary-structure-dependent quantities that matched closely with experimental values~\cite{RomanoOverstretching2013,Schreck2015}. By combining oxDNA molecular dynamics (MD) simulations of the aforementioned force clamp with our method for calculating internal forces, we demonstrate the importance of simulation-aided force calibration for DNA nanotechnological applications.

\section{Methods} \label{sec:methods}

\subsection{oxDNA Simulations}

The oxDNA model has been extensively described previously~\cite{Ouldridge2011,Sulc2012,Snodin2015}. Briefly, oxDNA represents nucleotides as rigid bodies featuring three interaction sites for backbone, stacking, and hydrogen-bonding interactions. Parameters in oxDNA were obtained using a ``top-down'' approach: the model was fitted to experimental results for structural, thermodynamic, and elastic properties of DNA. Nucleotides interact via pairwise potentials chosen to incorporate salient physical interactions, and the full potential energy function for the most recent version of the model, oxDNA2, contains terms for backbone connectivity, stacking, hydrogen bonding, cross stacking (i.e.\ stacking between bases on opposite duplex strands), excluded volume, coaxial stacking (i.e.\ stacking between contiguous bases on different strands, as across a nick), and electrostatic (according to Debye-H{\"u}ckel theory) interactions. While the model can capture salt, temperature, and sequence-dependence, it cannot represent the detailed interactions of DNA with solvent ions, though this shortcoming is not expected to affect the current work significantly. One additional, important caveat is that oxDNA currently cannot represent non-canonical base pairing, which could potentially affect its ability to model ssDNA secondary structures accurately. We do not believe this limitation to be consequential to our results, however, as detailed further below.

We performed MD simulations with oxDNA, using the velocity Verlet algorithm to integrate Newton's equations of motion with an integration time step of 15.2\,\si{\femto\second}. Solvent in oxDNA is implicit, and we simulated our systems in the canonical \emph{NVT} ensemble. Coupling between the system of interest and a heat bath was imitated using an Andersen-like thermostat~\cite{Andersen1980,Russo2009}, where velocities were redrawn from a Maxwell-Boltzmann distribution every 1.55\,\si{\pico\second}. The particle translational diffusion coefficient was set to 2.5, or $6.0\times10^{-7}$\,\si{\metre\squared\per\second} -- which is about $\sim$2 orders of magnitude faster than in experiment~\cite{Sulc_thesis} -- to accelerate diffusive dynamics and improve sampling. This change in the effective diffusion coefficient does not affect the equilibrium properties that we are calculating in this paper~\cite{FrenkelBookCh6}. For the simulations of the full origami structure shown in Figure~\ref{fig:fullnickels}, GPUs were used; otherwise, simulations were performed on CPU.

Before collecting data for analysis, all systems simulated in this work were equilibrated until the potential energy reached a stable value. Simulations were performed at 21\,\si{\celsius} and at varying monovalent salt concentrations, as detailed below. Complete simulation parameters used can be found in Table S1 in the Supporting Information.

\subsection{Calculating forces}
\label{sec:calcforce}
\begin{figure}
\begin{center}
\includegraphics[width=0.85\textwidth]{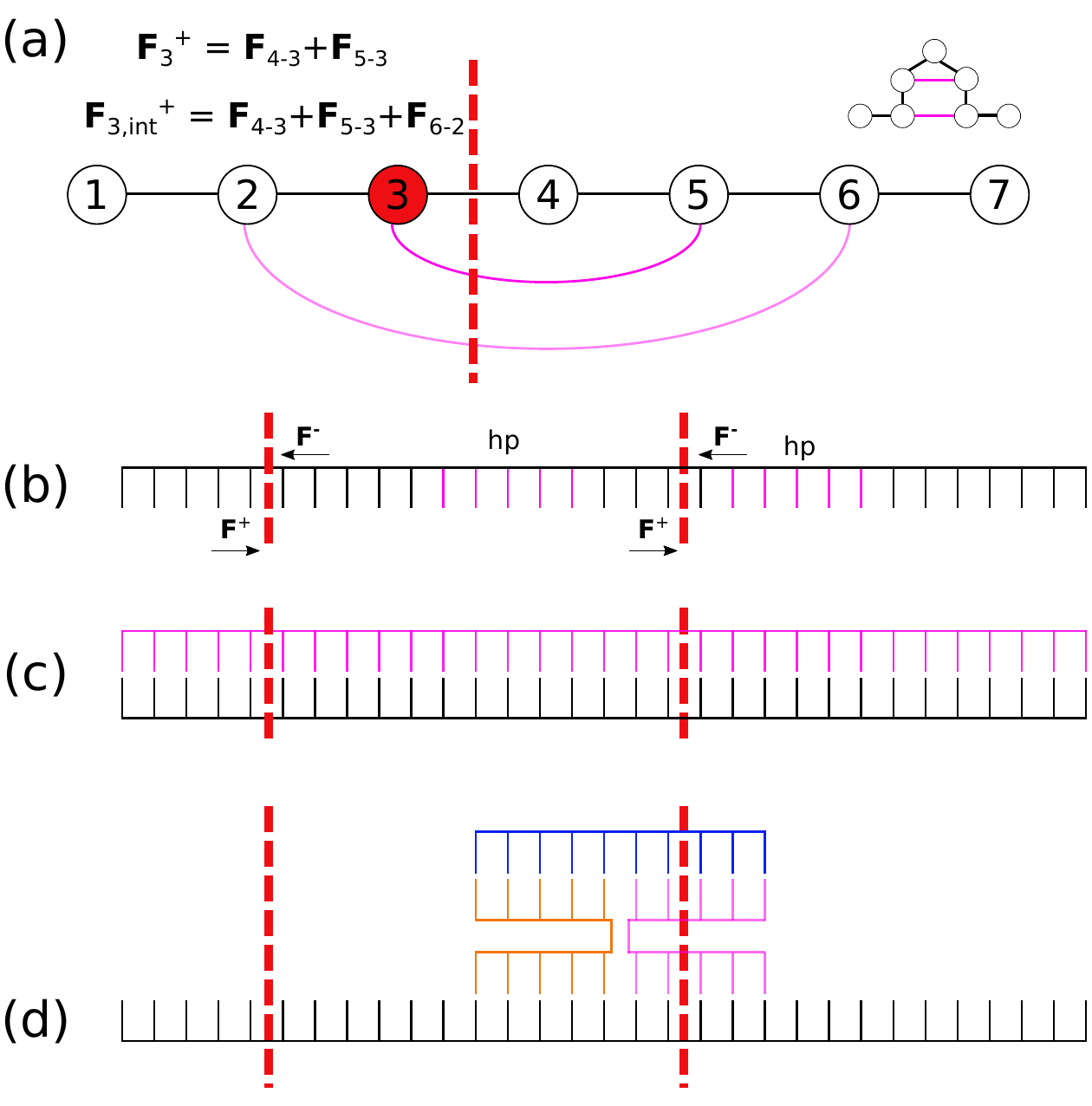}
\caption[]{(a) Schematic of the method for determining internal tension in DNA structures. For simplicity, consider nucleotides linked only through a FENE backbone (straight lines) and hydrogen bonds (pink, curved lines). The ssDNA depicted contains a 2 nucleotide hairpin stem with a 1 nucleotide loop, `unrolled' from the configuration shown in the upper right. Calculating the one-sided force on nucleotide 3 from topologically upstream yields $\mathbf{F}_{3}^+ = \mathbf{F}_{4-3}+\mathbf{F}_{5-3}$. The net force thus computed neglects an important counterbalancing term and yields too-large forces (Fig.~\ref{fig:ssDNA_example2}(b)) in the vicinity of secondary structure. Instead, the total force exerted on \emph{all} nucleotides to the left of a virtual interface (red dashed line) by \emph{all} nucleotides to the right of that interface should be computed: $\mathbf{F}_\text{3,int}^+ = \mathbf{F}_{4-3}+\mathbf{F}_{5-3}+\mathbf{F}_{6-2}$. The method is illustrated for three representative DNA structures: (b) a single strand with a hairpin (pink, indicated with `hp'); (c) a double strand; (d) a single strand containing a Holliday junction. Red dashed lines are interfaces across which the net force is calculated.}
\label{fig:top_int_schematic}
\end{center}
\end{figure}

In order to calculate the internal tension in a polymer using a coarse-grained simulator such as oxDNA, some average over the pairwise particle forces must be taken. Because nucleotides can interact with more than just their nearest neighbours, the correct way to find the average force at any point in the chain is to draw a `topological interface' between two nucleotides in the DNA structure (for example, the red dashed lines in Figure~\ref{fig:top_int_schematic}) and sum forces exerted by all nucleotides \emph{topologically downstream} of the interface on all nucleotides \emph{topologically upstream} of the interface. On average and at equilibrium, the force directed topologically upstream $\rightarrow$ downstream will be equal in magnitude to the force directed topologically downstream $\rightarrow$ upstream, and will be the net tension in the chain. In this and subsequent sections, we refer to the force calculated in one topological direction (upstream $\rightarrow$ downstream or vice versa) the `one-sided' force. In principle, given sufficient sampling, the one-sided force will be the same regardless of where in the DNA structure one draws the interface. Note that the sum of oppositely-directed one-sided forces at any interface should give zero at equilibrium for a polymer at rest.

To see this more clearly, consider the ssDNA schematic in Figure~\ref{fig:top_int_schematic}(a) of a single strand featuring a small hairpin. For simplicity, consider only the backbone interactions -- modelled in oxDNA by a finitely extensible nonlinear elastic (FENE) potential and indicated by straight lines -- and hydrogen bonding interactions -- indicated by the pink curved lines. A naive approach to calculating the tension in the strand connecting nucleotides 3 and 4, for instance, is to find the total one-sided force on nucleotide 3 from `upstream': the sum of the forces exerted on it by nucleotides 4 and 5, $\mathbf{F}_3^+$ in Fig.~\ref{fig:top_int_schematic}(a). This general strategy yields inconsistent forces across the polymer, with very large forces in regions of secondary structure, apparent in Figure~\ref{fig:ssDNA_example2}(b). Because we expect a constant one-sided tension throughout the polymer chain at equilibrium, it is clear that a different approach is needed. In the case of the example of Figure~\ref{fig:top_int_schematic}(a), the force exerted by nucleotide 6 on nucleotide 2 by virtue of their hydrogen bond must also be included, to correctly counterbalance the complex force propagation dynamics in the hairpin; this yields $\mathbf{F}_{\text{3,int}}^+$ in Fig.~\ref{fig:top_int_schematic}(a). The example discussed in reference to Figure~\ref{fig:ssDNA_example2} below confirms that this is the correct approach where one desires to compute the spatially-constant net internal tension across a polymer chain. Figure~\ref{fig:top_int_schematic}(b)--(d) illustrate example topological interfaces for a single strand with a hairpin, a double strand, and a single strand containing a Holliday junction, respectively. Here and in subsequent sections, we use the notational convention that $\mathbf{F_i}^+$ ($\mathbf{F_i}^-$) indicates the one-sided force directed topologically downstream $\rightarrow$ upstream (topologically upstream $\rightarrow$ downstream) at a given nucleotide or interface location $i$ in the polymer. Note that the topological direction must be distinguished from the \emph{physical} direction of the force. We indicate components of the force along particular physical axes using subscripts; for example, $F_{i,x}^+$ is the $x$-component of the topologically downstream $\rightarrow$ upstream one-sided force on nucleotide $i$ (for calculations involving the naive method) or across interface $i$ (for calculations involving the topological interface method).

In the sections that follow, we find the internal strand tension for test cases of a poly-T strand, ssDNA with a hairpin, and the nanoscopic force clamp of \citet{Nickels2016} that was discussed above.

\section{Test systems}

\subsection{poly-T ssDNA}
\begin{figure}
\begin{center}
\includegraphics[height=0.7\textheight]{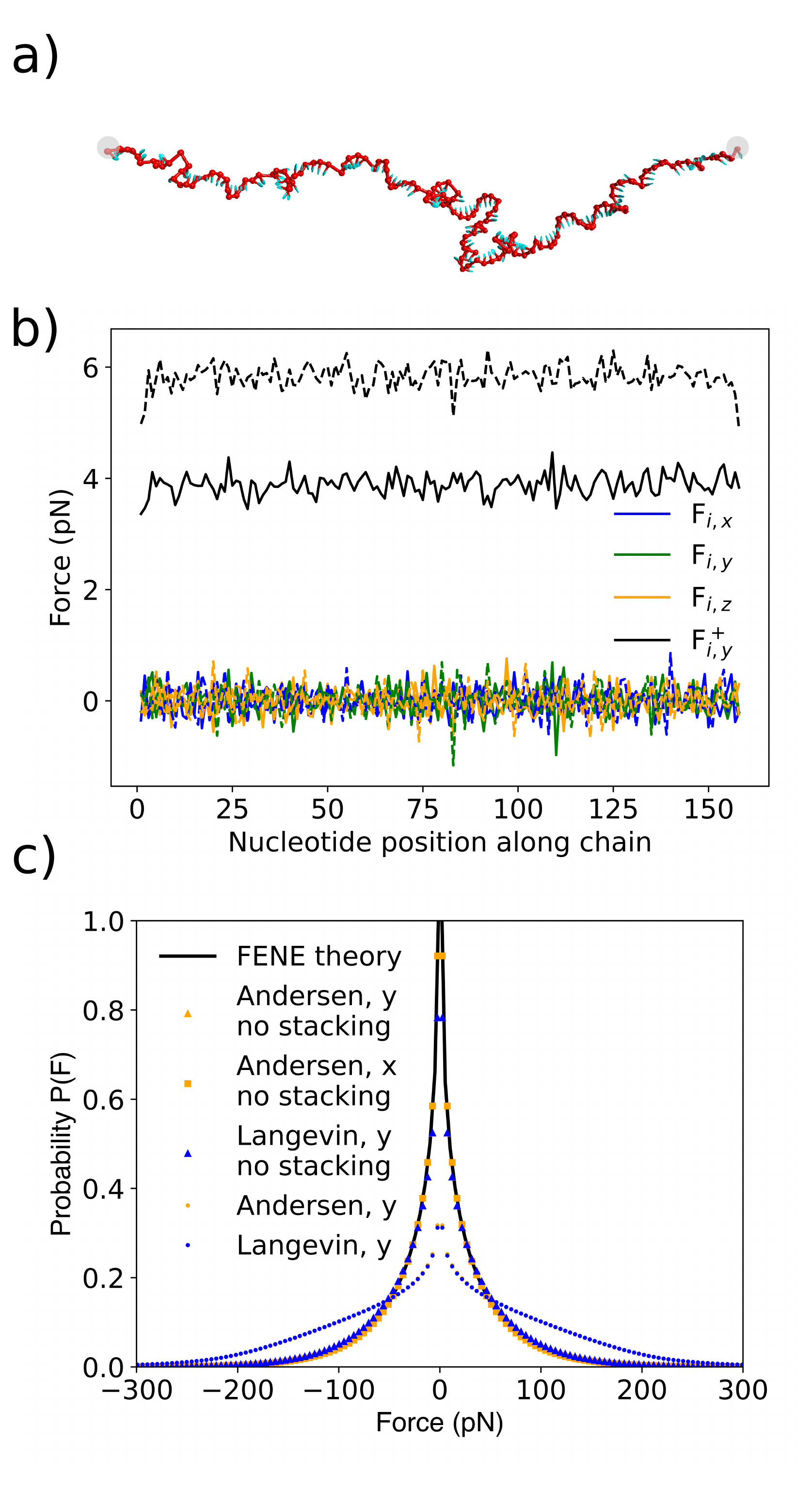}
\caption[]{(a) oxDNA representation of the 160-nt poly-T system, with end nucleotides subjected to harmonic traps (grey spheres). (b) Average forces in the single strand. Coloured lines represent the total force exerted on each nucleotide by all others, in all three directions. The black curves, labelled ‘$\text{F}_{i,y}^+$’, represent the ‘one-sided’ force on nucleotide $i$ along the direction of constraint, calculated using the naive method discussed above: the net force exerted on a nucleotide by all nucleotides topologically upstream of it. The dashed black curve is the result obtained when base stacking interactions in the strand are suppressed. (c) Force distribution averaged over all nucleotides in the strand of (a), both with the normal oxDNA model (circles) and with stacking interactions turned off (squares and triangles), using two different thermostats. Note that the results for the Andersen thermostat with no stacking (yellow triangles) are almost completely overlapped by the results for the Langevin thermostat with no stacking (blue triangles). Forces are net forces exerted on each nucleotide by all others, and are shown along the direction of constraint $y$ and along an orthogonal axis $x$. The black line is the theoretical expectation if the FENE potential is the only mediator of force.}
\label{fig:ssDNAexample1}
\end{center}
\end{figure}

Consider first a simple test case of a single 160-nucleotide (nt) poly-T strand whose ends are trapped in harmonic potentials, shown in Figure~\ref{fig:ssDNAexample1}(a). Constraining the endpoints should give rise to a nonzero tension in the strand. In this case, 
a naive attempt to compute this tension by summing the forces on a given nucleotide exerted by all other nucleotides topologically upstream of it yields similar values all along the chain: Figure~\ref{fig:ssDNAexample1}(b) displays in black the component of the one-sided force along the axis of constraint, here the y-axis, labelled `$F_{i,y}^+$', and equal to 3.84\,\si{\pn} when averaged over all 160 nucleotides. The average one-sided force component along each of the axes perpendicular to the axis of constraint is $\sim$0\,\si{\pn}. Additionally, the \emph{net} force on each nucleotide ($\mathbf{F_i}^{\text{net}} = \mathbf{F_i}^+ + \mathbf{F_i}^-$) averages to $\sim$0\,\si{\pn}, as as expected, given that the chain is in equilibrium and therefore is experiencing no net acceleration. Simulations were repeated with the base stacking interaction turned off; the resulting average force of $\sim$5.79\,\si{\pn}, shown by the dashed black line in Figure~\ref{fig:ssDNAexample1}(b), is higher than with stacking turned on. This is physically reasonable: a stacked chain has a longer Kuhn length than an unstacked one, and has fewer phase space conformations available to it; thus, its internal tension is lower.

We can compare this force to the prediction of the exFJC model:

\begin{equation}
\langle x \rangle = L_c\Bigg(\coth{\Big(\frac{F\ell}{k_BT}\Big)}-\frac{k_BT}{F\ell}\Bigg)\Bigg(1 + \frac{F}{S}\Bigg),
\label{eq:exFJC}
\end{equation}

\noindent where $F$ and $\langle x \rangle$ are the force and average extension, respectively; $S$ is a stretching modulus; $L_c=N\ell$ is the contour length of the chain; and $k_BT$ is the usual thermal energy. Under our simulation conditions, and using the frequently-cited parameters reported by \citet{Smith1995}: Kuhn length $\ell=1.5$\,\si{\nm}, contour length $L_c=$160 nt $\times$ 0.56\,\si{\nm}, and elastic modulus $S=800$\,\si{\pn}, the exFJC model predicts an entropic tension in the chain of $\sim$4.4\,\si{\pn}. That this is higher than the simulated result reflects that the experimental fits of \citet{Smith1995} are averaging over secondary structure, which is not present in the poly-T.

We note that the effective force shown in Fig.~\ref{fig:ssDNAexample1}(b) is an average, and that instantaneous forces can be much larger. We can consider this in more detail by looking at the distribution of forces experienced by an individual nucleotide along the chain, shown in Fig.~\ref{fig:ssDNAexample1}(c) (averaged over all 160 nucleotides in the chain) for two directions: along the axis of constraint ($y$) and orthogonal to the axis of constraint ($x$), and for two different thermostats: Andersen-like and Langevin (described on pages 1-2 in the Supporting Information). Data for the $x$ and $y$ axes are in good agreement, as are data for the two thermostats, which confirms that the distributions are not artefacts of a particular thermostat choice. The first thing to notice is that the distributions are very broad, with instantaneous forces over 100\,\si{\pn} in magnitude being common; therefore, averages must be computed over lots of data to derive sensible average force estimates (the values in Figure~\ref{fig:ssDNAexample1}(b) represent averages over 3.8$\times10^{10}$ MD timesteps, or $\sim$0.6~\si{\milli\second}.\footnote{In oxDNA, simulation time units can be converted to physical units based on simulation energy and length scales; this conversion yields 3.03$\times$10$^{-12}$\si{\second} per time unit. The simulations performed in this work used 200 MD steps per time unit, corresponding to $\sim$15~\si{\femto\second} per MD step. However, comparison of coarse-grained MD timescales to actual physical timescales must be done cautiously, due to the fact that coarse-graining reduces timescale separations between different processes~\cite{Padding2006}. This can be done in general by comparing the physical (experimental) and simulation timescales for a relevant process~\cite{Snodin2016}.}). Secondly, they are non-Gaussian. Finally, force distributions corresponding to the full simulations (circles) are broader than those with stacking turned off (triangles and squares). In both cases, the considerable magnitude of the forces and the shapes of the distributions can be rationalised by looking in detail at the underlying oxDNA potentials. 

For a single poly-T strand of DNA, the primary mechanisms available for internucleotide force transfer are propagation along the backbone, mediated by the FENE potential, and stacking interactions between bases. The force probability distribution expected if interactions are due solely to the FENE potential can be derived; see pages 2-4 of the Supporting Information for details. The resulting $P(F)$ is plotted in black in Fig.~\ref{fig:ssDNAexample1}(c), and matches well the data from simulations with stacking interactions suppressed. 

To discover why the stacking interaction broadens the force distributions, we can make a simple argument based on the form of the stacking potential in oxDNA. A rough estimate of representative forces at equilibrium due to each of the FENE and stacking potentials is given by the derivatives of these potentials (considering only the radial term of the stacking potential for simplicity) evaluated at the thermal energy that the system is expected to possess in equilibrium according to equipartition, $k_BT/2$; see pages 4-6 of the Supporting Information for details. Using oxDNA parameters, representative radially-directed forces are are $F^{thermal}_{FENE} \sim 89$\,\si{\pn} and $F^{thermal}_{stack} \sim 193$\,\si{\pn}, which, after taking into account that only one component of the force is plotted in Fig.~\ref{fig:ssDNAexample1}(c), match the distribution widths well.

\subsection{ssDNA with hairpin}

\begin{figure}
\begin{center}
\includegraphics[height=0.75\textheight]{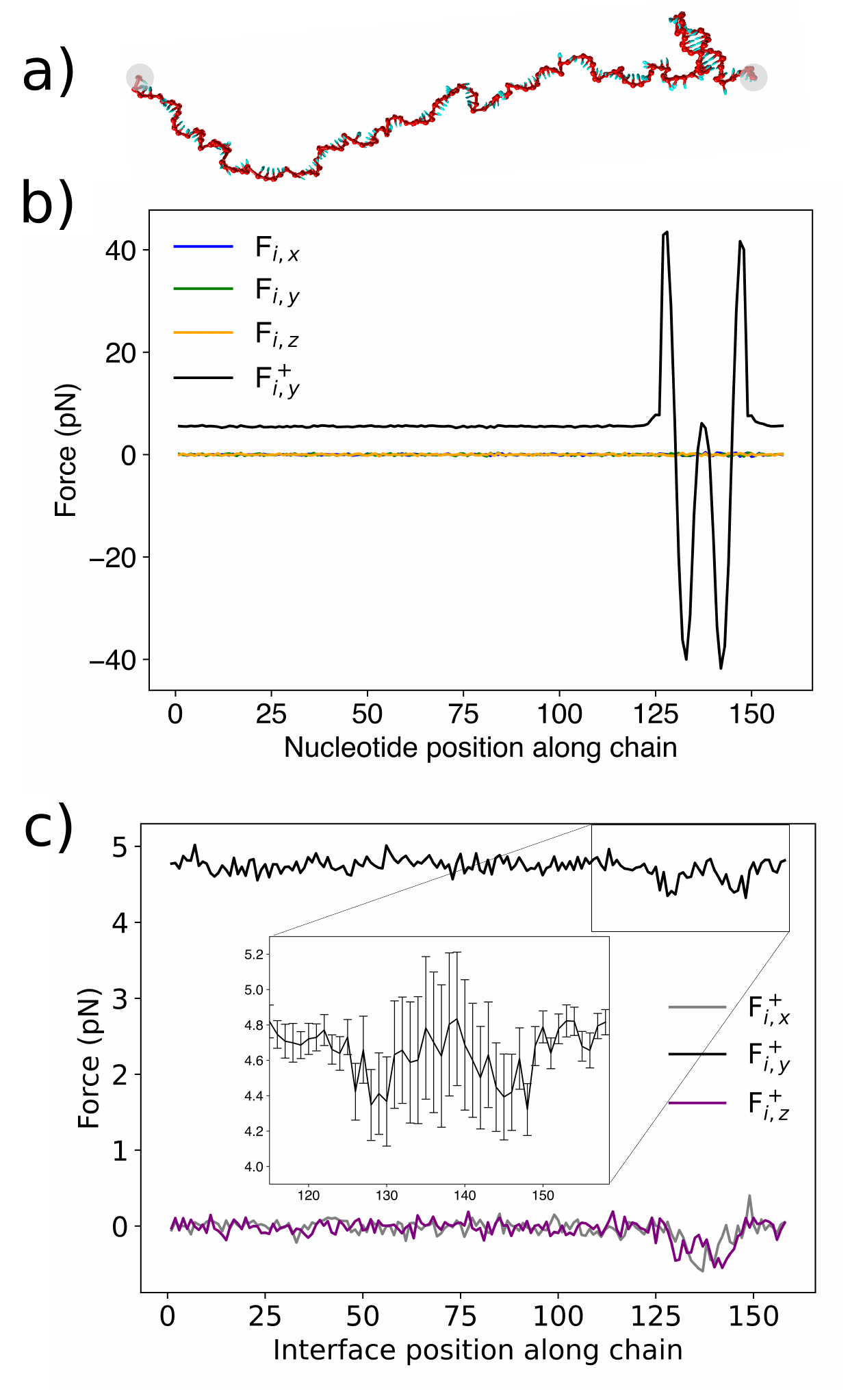}
\caption[]{(a) oxDNA representation of a strand identical to that of Fig.~\ref{fig:top_int_schematic}(a), but with a hairpin with a 10 bp-stem and a 4-T loop near its edge. (b) The net forces exerted on each nucleotide by all other nucleotides in three dimensions (blue, green yellow), as well as the one-sided force -- the net force exerted on each nucleotide by all nucleotides topologically upstream of it -- along the axis of constraint (black, $F_{i,y}^+$). The presence of secondary structure complicates force propagation in the strand. (c) One-sided forces computed instead using the topological interface method described in the text. Here, interface position $i$ indicates an interface drawn immediately topologically upstream of nucleotide $i$, as in Figure~\ref{fig:top_int_schematic}. Inset: zoom in of the region containing a 10-base pair hairpin. Error bars are standard errors of the mean of 64 identical parallel simulations, and tend to be larger in the vicinity of the hairpin because of the larger magnitude of forces in that region. As expected, a non-zero tension exists along the direction of constraint ($F_{i,y}^+$).}
\label{fig:ssDNA_example2}
\end{center}
\end{figure}

Consider now a more complex arrangement: a 160-nt strand that is poly-T save for a hairpin with a 10-bp stem and a 4-T loop near to the 5' end of the strand; see Figure~\ref{fig:ssDNA_example2}(a). Applying the method of section \ref{sec:calcforce} to calculate internal tension no longer works, as revealed by Figure~\ref{fig:ssDNA_example2}(b). The net force exerted on each nucleotide is still $0\,\si{\pn}$ on average, as it should be. However, while one expects the average one-sided force on each nucleotide along the direction of constraint to be approximately constant and equal to the entropic tension in the strand, this is not the case in the vicinity of the hairpin. Clearly, when analyzing tension in a DNA chain when secondary structure is involved, it is no longer sufficient to simply calculate the forces exerted on each nucleotide by all nucleotides topologically `upstream'  or `downstream' of it, as force propagation in that case is mediated by a more complex network of interactions -- including hydrogen bonding and cross-stacking -- that communicate tension through the chain in a non-trivial manner. That the deviations in $F_{i,y}^+$ forces in the hairpin region are so large partly reflects the features of the oxDNA potential, which in turn are capturing the physical frustration inherent to DNA interactions. In particular, bases in the hairpin stem cannot simultaneously minimize hydrogen-bonding, stacking, FENE, and cross-stacking potentials, regardless of whether there is a net tension in the strand; this is evidenced by the fact that, for example, a stacked single strand has a tighter helix than dsDNA~\cite{Chen2007} and that base-paired nucleotides have a small propeller twist~\cite{CalladineBook2004}. Because they are displaced from the minima of one of more of these potentials, the average individual inter-nucleotide forces in the vicinity of the hairpin can be large; hence, when only a subset of the forces across a topological interface are taken into account, there will be large deviations from the true force being transmitted through the chain. 

The approach described above in \emph{Methods}, however -- adding \emph{all} forces directed topologically upstream $\rightarrow$ downstream across a given interface -- recovers the expected results: see Fig.~\ref{fig:ssDNA_example2}(c). In this case, the tension averaged across all nucleotides along the axis of constraint is $\sim4.8$\,\si{\pn} (higher than in the poly-T case previously considered because the hairpin has shortened the effective contour length), and the force behaves reasonably in the vicinity of the hairpin, though errors in that region are larger due to the greater magnitude of pairwise nucleotide forces.

\subsection{Nanoscopic force clamp}
\begin{figure}
\begin{center}
\includegraphics[width=\textwidth]{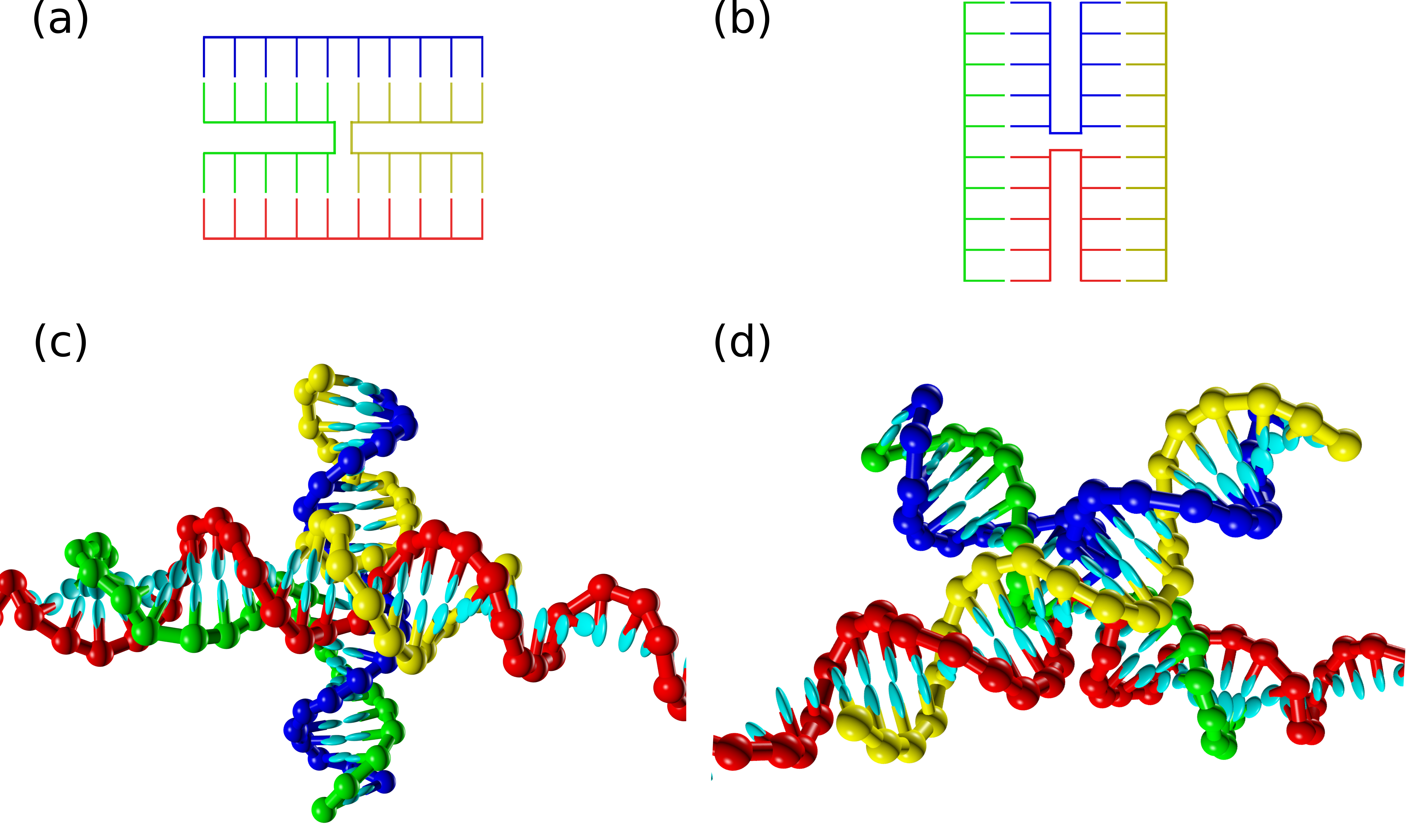}
\caption[]{Schematics and oxDNA representations of the two possible coaxially-stacked isomers of the Holliday junction. (a), (c) IsoII, in which coaxial stacking occurs between yellow and green strands. (b), (d) IsoI, in which coaxial stacking occurs between red and blue strands. Here, the red strand is part of a longer ssDNA segment (see Fig.~\ref{fig:HJ_inner_regions} for zoom-out) along which force is applied. This force adds a thermodynamic bias favouring IsoII.}
\label{fig:HJ_isomers}
\end{center}
\end{figure}

\begin{figure}
\begin{center}
\includegraphics[width=0.75\textwidth]{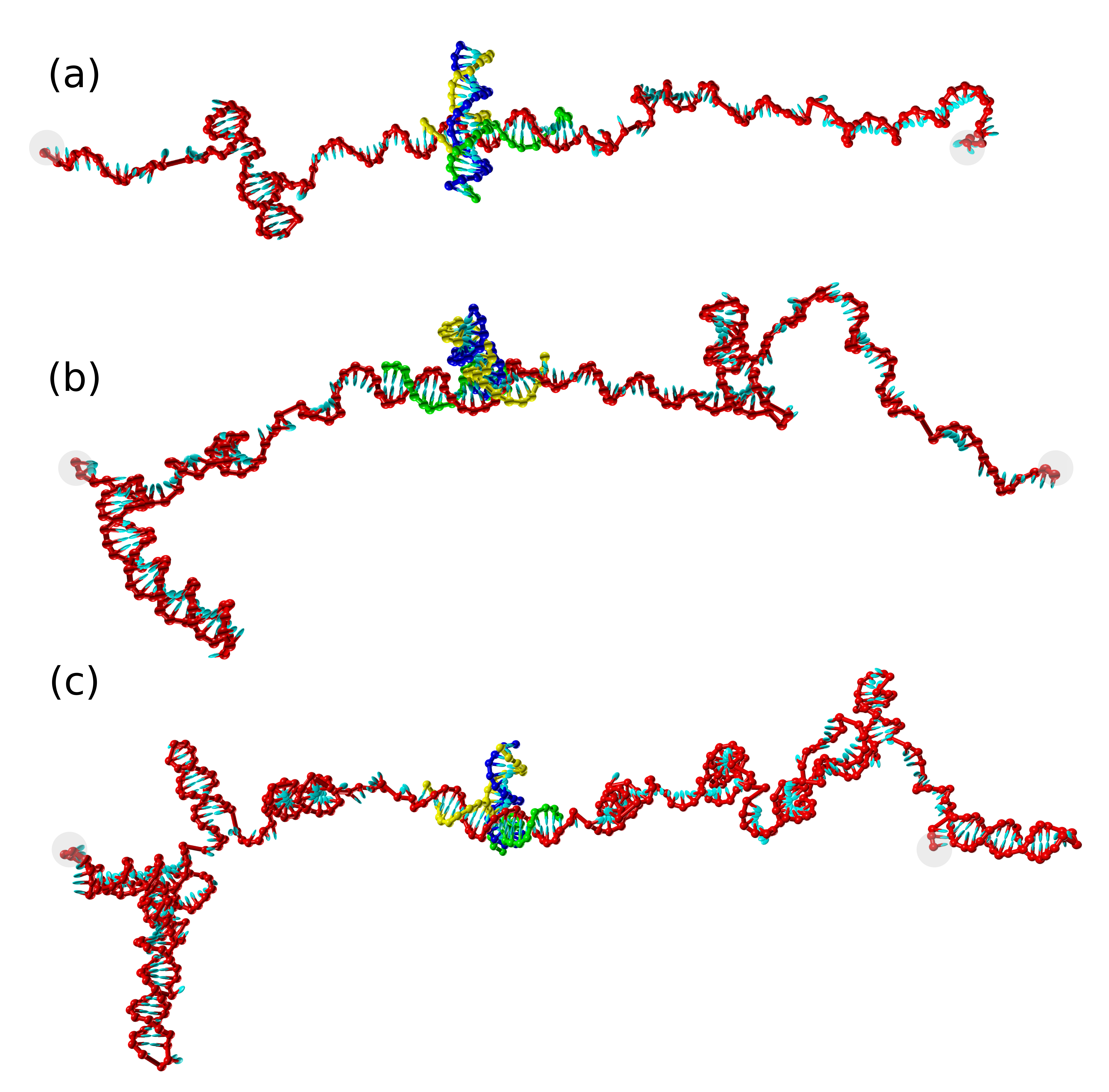}
\caption[]{oxDNA representations of the gap-spanning ssDNA segment of the force clamp, with Holliday junction attached, for the (a) 4.0\,\si{\pn}, (b) 2.5\,\si{\pn}, and (c) 1.2\,\si{\pn} designs. The red strand contains a total of 162-nt, 228-nt, and 424-nt in each of these three cases, respectively. Clearly, this sequence is prone to secondary structure formation, which is more prevalent at lower forces, as expected.}
\label{fig:HJ_inner_regions}
\end{center}
\end{figure}

With some improved understanding of the mechanisms of force propagation in oxDNA simulations and the appropriate methods for calculating internal forces in hand, we turn our attention to the more interesting case of the nanoscopic force clamp of \citet{Nickels2016}, described in the \emph{Introduction}. 

Their design solves many of the issues with traditional force spectroscopy apparatus like atomic force microscopes and laser optical tweezers, and is a powerful addition to the toolbox used to investigate mechanical processes in the cell.  In particular, because it is instrument-free, many of the artefacts that plague single molecule force spectroscopy (SMFS) experiments -- for example, convolution with long linker handles~\cite{Woodside2014} -- are not a concern. Additionally, the nanostructure is highly scalable, so single-molecule experiments can be carried out in parallel in large numbers; this can be contrasted with the fundamentally serial nature -- and consequently poorer statistics -- of other experimental SMFS strategies. Finally, the need for attachment to microscopes prevents traditional SMFS experiments from being performed \textit{in vivo}, while no such limitation exists in theory for nanoscopic force clamps. 

\citeauthor{Nickels2016} used their force clamp to study the Holliday junction~\cite{Hohng2007} and the tension-dependent binding of the TATA protein~\cite{Nickels2016}. The precise forces exerted on the systems of interest were calculated as follows:
\begin{itemize}
\item the origami was assumed to be completely rigid;
\item the ssDNA was replaced with an exFJC polymer with end-to-end distance given by the designed gap size minus the average extension of the molecule of interest;
\item parameters for Kuhn length, contour length, and elastic modulus derived from previous experiments were plugged into the exFJC force-extension equation and combined with the end-to-end distance estimates to yield forces.
\end{itemize}

Three potential problems with this procedure, elements of which many DNA tension probe studies share~\cite{Liu2016,Brockman2018,Ma2019,Liedl2010,Zhang2014,Blakely2014,Kramm2020}, bear consideration. Firstly,
while dsDNA is well-described by the WLC\cite{Marko1995,Gross2011,Nomidis2017}, 
ssDNA behaviour is difficult to capture with simplified polymer models. Authors have used both WLC and exFJC to fit ssDNA in different regimes, dependent on ionic conditions and force~\cite{Huguet2010,Ritort2014a}. At low ionic concentrations, electrostatic repulsion fosters increased rigidity of ssDNA, so the WLC model may be preferable to the exFJC, which works better at higher salt concentrations~\cite{Huguet2010}. Both models, however, neglect excluded volume and stacking interactions between ssDNA bases. 
Difficulties inherent in mapping ssDNA to a WLC or exFJC emerge when comparing elastic parameters obtained from experiments performed at zero force (where bases are free to stack) and force-pulling experiments (featuring higher forces at which stacking is less favourable) that feature a clear stacked $\rightarrow$ unstacked transition~\cite{Mcintosh2014,Ritort2014a}. A dependence of fit parameter on experimental regime is widespread in the literature~\cite{Smith1995,Weiss2005,Ritort2014a,Saleh2017}, but interestingly this has not been extensively acknowledged.

A second issue with force probe calibration using polymer models is the existence of sequence-dependent secondary structures in ssDNA, which reduce the effective contour length of ssDNA and elevate internal forces. One of the most widely-used sets of ssDNA elastic parameters is that derived by ~\citet{Smith1995}, who fitted force-extension curves of $\lambda$-DNA to the exFJC. These parameters represent an average over $\lambda$-DNA secondary structure and are thus not transferrable to different ssDNA sequences.

Finally, assumptions about the rigidity of the origami portion of the force clamp could introduce errors into the force estimates. DNA origami, of course, have some degree of flexibility~\cite{Lei2018}, but the elastic mechanical properties have only been characterized for a limited number of rod-like origami.\cite{Kauert2011,Pfitzner2013}

As mentioned, \citet{Nickels2016} demonstrate the operation of their nanoscopic force clamp by studying the four-way Holliday junction~\cite{Hohng2007}(HJ), a fundamental component of DNA origami designs. HJs are comprised of 4 strands hydrogen-bonded to each other and arranged in an `X' shape. In the presence of salt, the strands coaxially stack (i.e.\ bases belonging to two separate strands stack with one another) to adopt one of two possible isomeric conformations, illustrated in Figure~\ref{fig:HJ_isomers}. In Fig.~\ref{fig:HJ_isomers}(a) and (c), the green and yellow strands are coaxially stacked, and the blue and red strands are regularly stacked such that they are straight; the reverse is true in Fig.~\ref{fig:HJ_isomers}(b) and (d). The population ratio between the isomeric states can be altered via the application of external force; pulling on either end of the red strand in Fig.~\ref{fig:HJ_isomers}, for example, will favour the isomer in which the red strand is straight. \citet{Nickels2016} refer to this more favourable isomeric state as `IsoII', and the less favourable isomer, in which the tension-bearing strand is bent, as `IsoI', drawing their nomenclature from the low (IsoI) and high (IsoII) FRET states for a particular geometry of donor/acceptor dyes on the chosen HJ sequence, following \citet{Hohng2007}. We adopt here the same convention.

To illustrate the force-dependence of the population ratios of IsoI and IsoII, \citet{Nickels2016} construct three versions of the origami force clamp. The designs were named according to the forces the authors calculate the ssDNA springs to exert on the molecule of interest: 4.0\,\si{\pn}, 2.5\,\si{\pn}, and 1.2\,\si{\pn}. Values for these designed forces were derived from the exFJC model, Eq.~\ref{eq:exFJC}, using the Kuhn length and stretch modulus parameters reported by \citet{Smith1995}; the number of nucleotides spanning the central gap in each design (160, 4.0\,\si{\pn}; 226, 2.5\,\si{\pn}; 422, 1.2\,\si{\pn}) minus the 22 nucleotides corresponding to the HJ region; a contour length per base pair of 0.63\,\si{\nm}, derived from crystallographic experiments~\cite{Murphy2004}; and the designed central gap width of 42.84\,\si{\nm} -- labelled $d$ in Fig.~\ref{fig:fullnickels} -- minus the average end-to-end extension of the HJ as the polymer end-to-end distance, $\langle x \rangle$. 
Notably, the magnesium ion concentration at which the force clamp experiments were performed -- 100\,\si{\milli\Molar} [Mg$^{2+}$] -- represents much higher electrostatic screening than the salt concentration used in the experiments of \citet{Smith1995}, 150\,\si{\milli\Molar} [Na$^{+}$], because Mg$^{2+}$ has been estimated to have an effect that is 20-100$\times$ stronger than Na$^{+}$~\cite{Dietz2012a,Ritort2014a}, depending on the property that is being studied. High ionic strengths are often used in DNA origami experiments to facilitate origami assembly.

\subsubsection{Simulation strategy}

For our purposes of determining the internal forces in the single-stranded section of the origami force clamp, it was not necessary to simulate the full origami structure in the force-measurement simulations. Instead, we simulated only the ssDNA strand in the central gap of the force clamp and represented the constraints placed on the single strand by the origami with harmonic traps (where the stiffness and separation of the traps were extracted from simulations of the full origami), thus vastly improving the statistics we were able to collect in our available simulation time. To achieve this, we first performed an MD simulation of the origami force clamp with no connecting single strand, and monitored the end-to-end distance $r$ between scaffold nucleotides on either end of the clamp; the results are shown in Figure S2. The resulting distribution was fit to a Gaussian form:
\begin{equation}
P(r) = Ae^{-\frac{(r-\langle r \rangle)^2}{2\sigma^2}}
\end{equation}
\noindent to extract the mean $\langle r \rangle = 42.77$\,\si{\nm} and variance $\sigma_r^2 = 0.50$\,\si{\nano\metre\squared}. Note that $\langle r \rangle$ is approximately equal to the value used by \citet{Nickels2016} based on simple geometrical arguments: 42.84\,\si{\nm}. The values for $\langle r \rangle$ and $\langle \sigma_r^2 \rangle$ can be reproduced by two 3-dimensional harmonic traps of stiffness $k=16.32$\,\si{\pnnm} and trap separation $d=42.76$\,\si{\nm} acting on the end nucleotides of the tension-bearing single strand (the approach used to obtain these values is detailed on pages 6-7 of the Supporting Information). We note that by only simulating the force-bearing strand, we are ignoring potential interactions between this section and the rest of the origami (aside from these 3D harmonic traps on its ends). These interactions will be predominantly due to the excluded volume, but given the design (e.g. the clearance above bridging section) we don't expect their effects to be significant.

We performed MD simulations on the ssDNA + HJ system, with end nucleotides harmonically trapped, for the three different force designs of \citet{Nickels2016}; see Figure~\ref{fig:HJ_inner_regions}. Including the two trapped endpoint nucleotides, the main ssDNA strands comprised 162-nt, 228-nt, and 424-nt for the 4.0\,\si{\pn}, 2.5\,\si{\pn}, and 1.2\,\si{\pn} designs, respectively. Clearly, secondary structure is present for this ssDNA sequence, shortening the effective contour length of the ssDNA. We used the same temperature as \citeauthor{Nickels2016} and a monovalent salt concentration of [Na$^{+}$]=5\,\si{\Molar} in order to mimic their high experimental [Mg$^{2+}$] concentration. To explore the effect of origami compliance on the force in the structure (the exFJC model used by \citet{Nickels2016} assumes fixed endpoints), we also performed simulations on the 4.0\,\si{\pn} system using harmonic trap stiffnesses an order of magnitude larger and smaller than $k=16.32$\,\si{\pnnm}. Finally, to explore the effect of salt concentration on the results, we performed additional simulations on the 4.0\,\si{\pn} system at [Na$^{+}$]=0.15\,\si{\Molar}, 0.5\,\si{\Molar}, and 1.0\,\si{\Molar}. In all instances, we simulated the strands both with secondary structure and without allowing secondary structure to form. 
Transitions between HJ IsoI and IsoII conformations are very rare on the timescale of our simulations\footnote[2]{Over $\sim5\times10^{11}$ cumulative MD simulation steps ($\sim$7\,\si{\milli\second}), only 14 transitions were observed, and in each case, the HJ never reverted. This was reasonable given that the experimentally-observed transition times are on the order of tens of milliseconds.~\cite{Nickels2016}}, so we were unable to probe the effect of force on transition frequency, but we simulated each isomer separately and report results for both.

It is important to ensure that our simulations are sampling all relevant secondary structure, a potentially difficult task given both the possibility of the system becoming trapped in local minima for long stretches of simulation time and the large number of potential different secondary structures. To gauge whether we had sampled secondary structure sufficiently, we split our simulation output for each system in half and tracked which bases formed hydrogen bonds throughout the simulations (see Supporting Figures S3-S5). The hydrogen-bonding pattern looked similar enough between the two simulation halves to give us confidence that our secondary structure sampling is sufficient in the present case; see pages 8-11 of the Supporting Information for details. An additional consideration is oxDNA's inability to capture non-canonical base pairing, as mentioned in the \emph{Introduction}. To explore this further, we ran our sequences through the Quadruplex forming G-Rich Sequences (QGRS) Mapper~\cite{KikinQGRS2006} to determine whether G quartets can form in our system. For the 162-nt and 228-nt designs, no G quartets are expected. For the 424-nt design, however, five G-quartet motifs (each with 2 stacked G tetrads) are theoretically possible. The consistency of the trends observed in our results across all three systems, detailed below, give us confidence that this limitation is not significant to our central results, however.

\subsubsection{Results}

\begin{figure}
\begin{center}
\includegraphics[width=\textwidth]{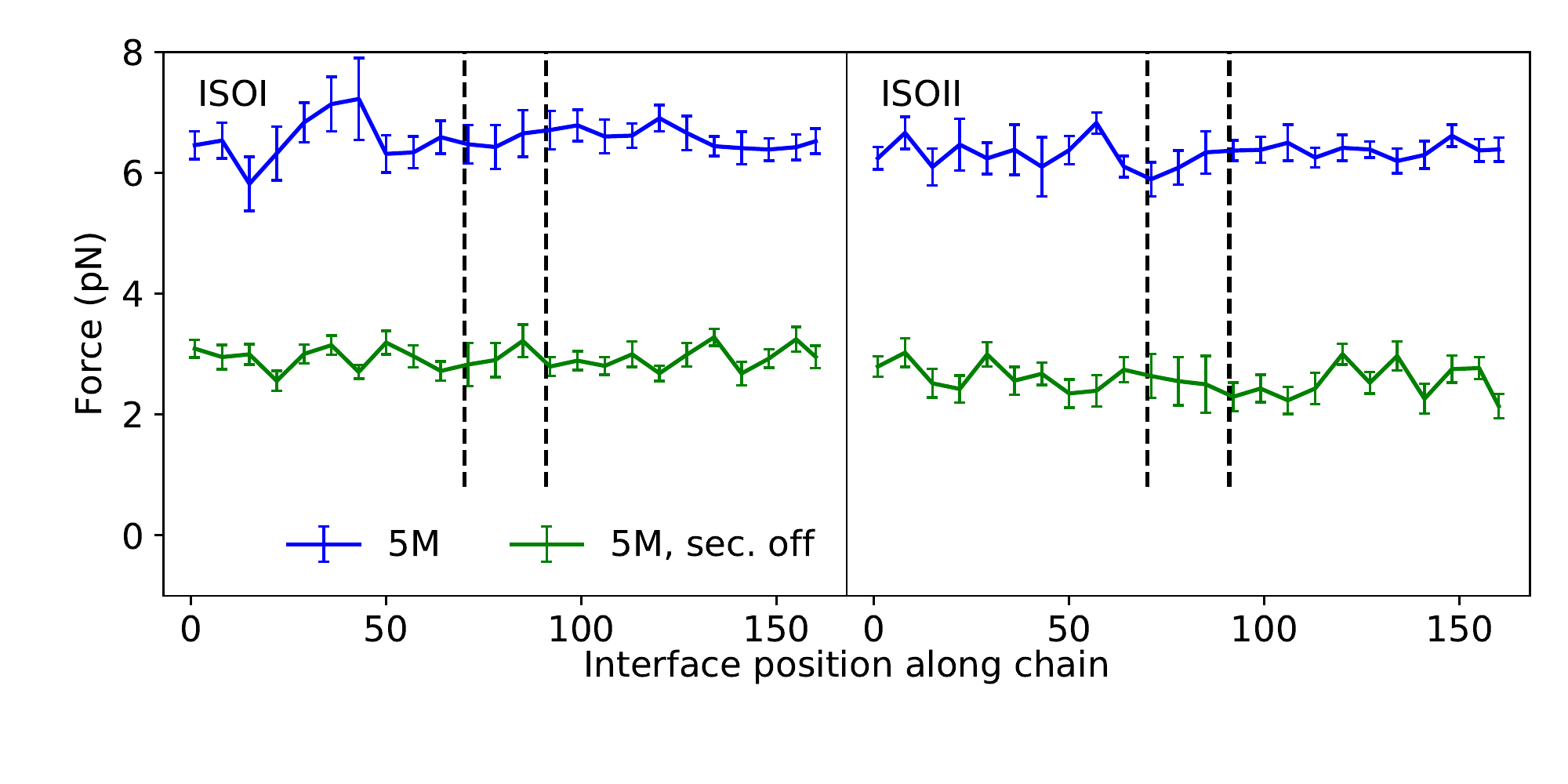}
\caption[]{One-sided forces crossing interfaces at 24 different locations along the ssDNA strand for \citet{Nickels2016}'s 4.0\,\si{\pn} design. The location of the HJ is denoted with black dashed lines. Simulations were performed at [Na$^{+}$]=5\,\si{\Molar} for IsoI (left) and IsoII (right), with secondary structure formation permitted (blue curves) and suppressed (green curves). Errors are standard deviations over 23 (left) or 24 (right) identical replica simulations of $\sim10^9$ MD steps ($\sim$35\,\si{\micro\s} (left) and $\sim$47\,\si{\micro\s} (right) total simulation time). 
}
\label{fig:force_v_nuc_compare}
\end{center}
\end{figure}
\begin{figure}
\begin{center}
\includegraphics[width=\textwidth]{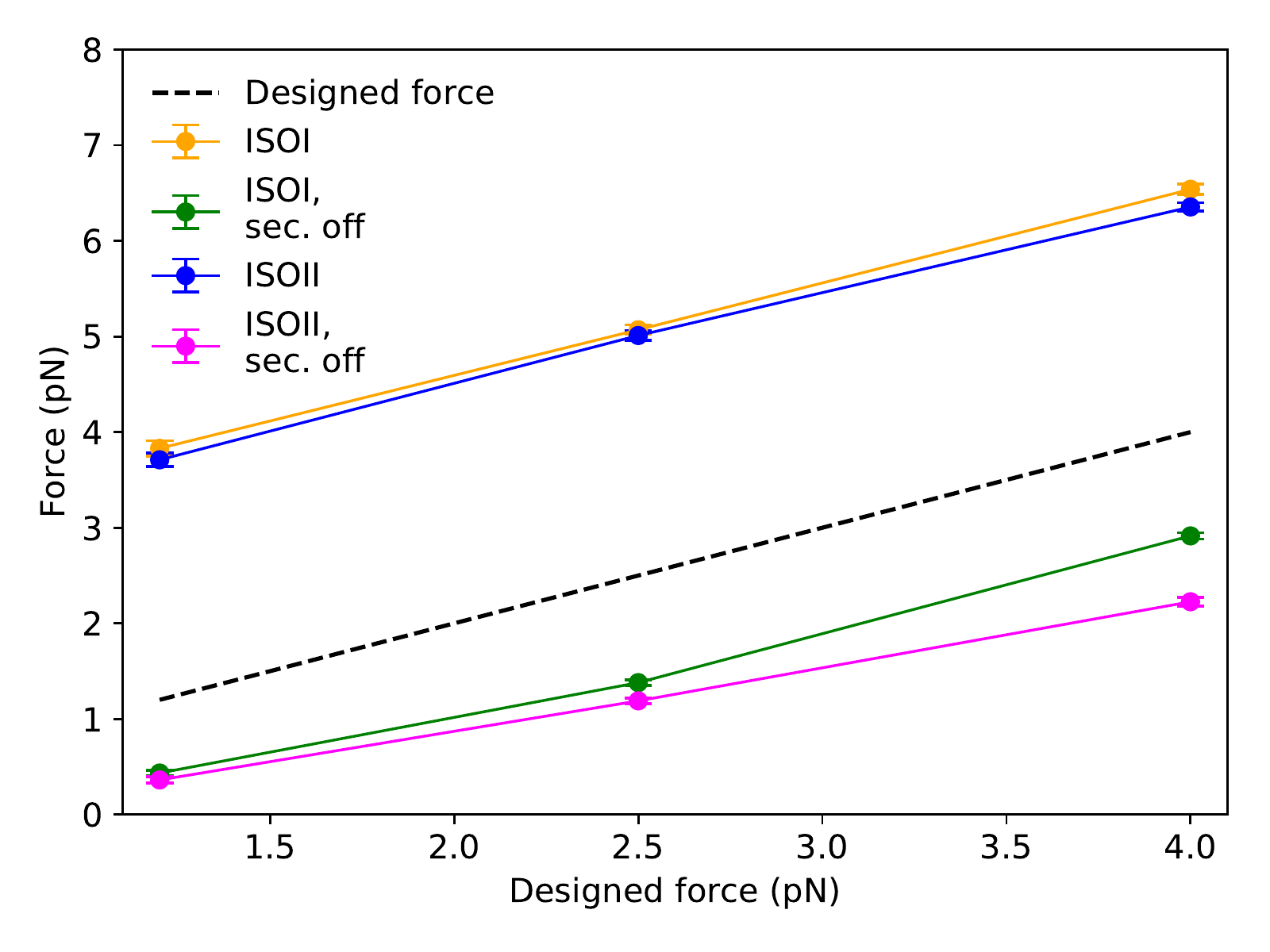}
\caption[]{Force measured in simulations versus `designed' force, based on the exFJC parameters of \citet{Smith1995} 1:1 equivalence between the two is indicated by the dashed black line. In all cases, the measured force is larger than the designed force (by around $\sim$2.5\,\si{\pn}) and larger than the result when secondary structure is suppressed (by around $\sim$3.6\,\si{\pn}). In all cases, the force difference between IsoI and IsoII variants is small.}
\label{fig:forcevforce}
\end{center}
\end{figure}
\begin{figure}
\begin{center}
\includegraphics[width=\textwidth]{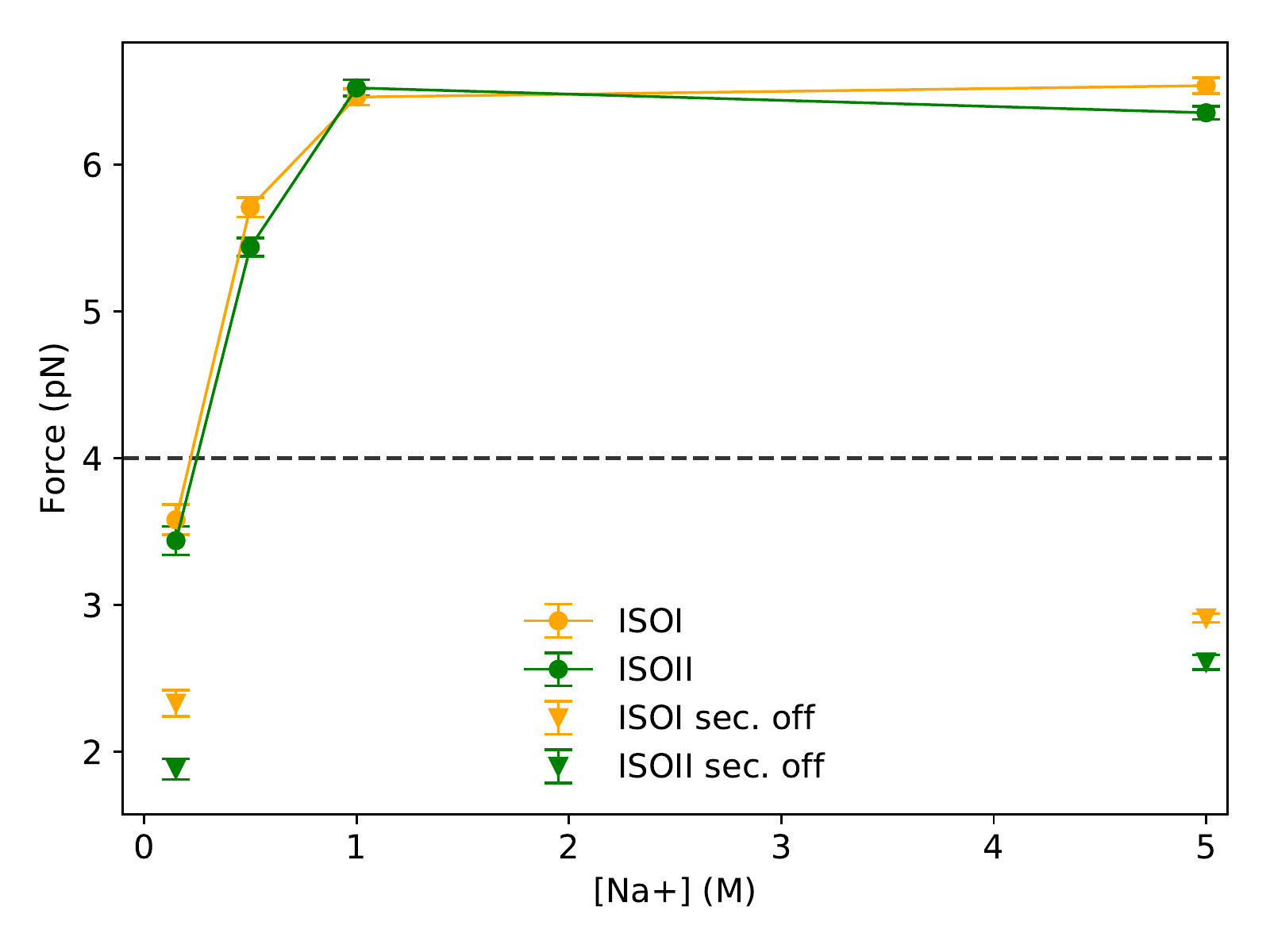}
\caption[]{Effect of salt concentration on internal force in ssDNA, for the 4.0\,\si{\pn} design of \citet{Nickels2016} (the designed force is indicated by the dashed black line and includes no salt dependence). At low salt, secondary structure is partially suppressed by electrostatic repulsion, leading to lower internal forces. This is corroborated by the comparatively smaller gap between low and high salt when secondary structure is suppressed.}
\label{fig:forcevsalt}
\end{center}
\end{figure}
The full results of our simulations are reported in Supporting Table S1. Figure~\ref{fig:force_v_nuc_compare} shows some representative one-sided forces crossing several interfaces along the ssDNA strand for the 4.0\,\si{\pn} design, where the interfaces are located immediately topologically upstream of the nucleotides indicated in the abscissa. Larger errors likely correlate with regions with an enhanced tendency to form secondary structure, because of the larger magnitude of the (positive and negative) forces across the interfaces in those regions; this effect was observed in Fig.~\ref{fig:ssDNA_example2}. Only the projection of the force onto the direction of the trap separation is shown; forces along the orthogonal directions average to zero, as expected.

Figures~\ref{fig:forcevforce}-\ref{fig:forcevsalt} contain results for forces averaged over 24, 33, and 39 different interfaces distributed along the strand for the 4.0\,\si{\pn} 2.5\,\si{\pn}, and 1.2\,\si{\pn} designs, respectively; error bars represent the standard error of a weighted mean. Immediately evident in Fig.~\ref{fig:forcevforce} is the substantial discrepancy between the forces as measured in simulation (blue and yellow curves) and both the \citet{Smith1995} exFJC force predictions (black dashed line; $\sim$2.5\,\si{\pn} average discrepancy) and the forces in the absence of secondary structure ($\sim$3.6\,\si{\pn} average discrepancy). Here, just as in the case of the simple 160-nt poly-T strand, forces in the absence of secondary structure are lower than the exFJC estimates; as aforementioned, this can be understood by considering that the fits performed by \citet{Smith1995} averaged over secondary structure inherent to $\lambda$-phage ssDNA.

While it is clear that secondary structure has a significant effect on internal force, another key driver of the disparity between measured forces and exFJC predictions in this case may simply be the different salt concentrations under which the experiments of \citet{Smith1995} and \citet{Nickels2016} were performed. Figure~\ref{fig:forcevsalt} shows the average forces for the 4.0\,\si{\pn} design at four different salt concentrations; at the lowest of these, [Na$^{+}$]=0.15\si{\Molar} -- equivalent to the concentration at which \citet{Smith1995} performed their fits -- the measured force is $\sim$3\,\si{\pn} lower than at the higher salt concentrations used by \citet{Nickels2016}, and much closer to the estimate of 4.0\,\si{\pn} obtained using \citet{Smith1995}'s parameters. Caution is clearly needed when translating ssDNA elastic fit results performed under different conditions. When secondary structure is suppressed, the difference in force between the highest and lowest ionic strengths is only $\sim$0.7\,\si{\pn}, underscoring secondary structure formation as the main driver of force variation with ion concentration. The remaining difference is attributable to the tendency of ssDNA to form more compact coils at higher salt concentration as increased screening reduces inter-phosphate repulsion~\cite{Mcintosh2014}.


We also explored the effect of the compliance of the attachment points on internal ssDNA forces; see Figure S7 in the Supporting Information for full details. In summary, while decreasing the trap stiffness by an order of magnitude (from $\sim$16.32\,\si{\pnnm} to $\sim$1.632\,\si{\pnnm}) decreases the force by about 0.2\,\si{\pn}, increasing the stiffness by an order of magnitude has no discernible effect. As oxDNA has been elsewhere shown to reproduce the elastic properties of DNA origami well~\cite{Sharma2017,Mishra2018}, we can conclude that for these designs the assumption of \citet{Nickels2016} that the origami is rigid does not affect the force estimate.   

As a final note, we point out that the difference in force between the two HJ isomers is in all cases rather small: $\sim$0.1-0.5\,\si{\pn}, with the IsoI conformation leading to higher (or the same within error) forces than IsoII, as expected. We do not expect to perfectly reproduce this difference, however, as the oxDNA model cannot fully capture the magnitude of the twist angle between the HJ arms~\cite{Snodin2019}.

\section{Conclusion}

We have demonstrated a method for calculating the internal forces in ssDNA to high statistical precision ($\sim$0.05\,\si{\pn}). While systematic errors intrinsic to the oxDNA coarse-grained model can be expected to affect the absolute accuracy of our force calculations, oxDNA has elsewhere captured the mechanical behaviour of ssDNA and dsDNA well~\cite{Ouldridge2011,Mosayebi2015,Nomidis2019}, which gives us confidence that our predictions are much more realistic than those of simplified polymer model fits. Further, given the limited precision of experimental calibration methods ($\sim$2-5\,\si{\pn}~\cite{Shroff2005,Dutta2018}), we argue that oxDNA predictions supply a viable alternative -- or at least complement -- to these approaches. 
In addition to its utility in computing forces in ssDNA-based force sensors, force clamp devices, and tensegrity nanostructures, our method can be expanded to calculate forces in more complex structures, such as DNA minicircles~\cite{Harris2016,Presern2018}. Furthermore, the method is general enough to be adapted to calculate forces in alternative coarse-grained models, including for non-DNA polymers like proteins.  

A few caveats should be noted. Firstly, the current work has assumed that origami structures were perfectly assembled -- with all staples present and no topologically incorrect strand routings -- despite the possibility of defects occurring experimentally which could compromise the effective stiffness of the origami or alter the length of the ssDNA spring. Furthermore, oxDNA has been parametrized to capture the [Na$^+$] dependence of secondary structure thermodynamics. To describe experiments featuring [Mg$^{2+}$], we relied on a mapping to an equivalent [Na$^+$]. In the current work, this was not problematic, as experiments were carried out in a high-salt regime where the predicted force had reached its limiting value, but for lower [Mg$^{2+}$], comparison of simulation and experiment may be less straightforward. Another consideration to bear in mind is the neglect of secondary structures involving non-canonical base pairs when calculating force averages. In systems where such motifs are expected to be relevant, deriving very accurate force estimates may necessitate the use of a coarse-grained model parameterized to treat non-Watson Crick base pairing. Lastly, it is possible that for high internal forces or low ionic strengths, the ssDNA tension may lead to non-elastic yielding events within origami designs. To capture such effects, the DNA origami would need to be simulated in its entirety.

Simulating precise experimental sequences under relevant conditions is crucial for obtaining accurate estimates of internal force, as using previously-obtained fit parameters with WLC and exFJC theoretical models can lead to incorrect estimates due to differences in secondary structure, stacking propensities, salt concentration, and stiffness of the endpoint traps, as well as fundamental deficiencies in the WLC and exFJC models, such as their neglect of excluded volume and base stacking. In particular, our results demonstrate that ignoring the difference between the experimental ionic solution conditions and those at which fit parameters were derived is a significant driver of errors in ssDNA force estimates for the nanoscopic force clamp of ~\citet{Nickels2016}. Future applications seeking to precisely determine internal forces of DNA, whether for the purposes of computationally modelling DNA elastic behaviour~\cite{Castro2011,Kim2012} or for understanding biomechanical systems~\cite{Kramm2020}, would benefit from the force estimates provided by oxDNA or similar coarse-grained models.

\begin{acknowledgement}

MCE thanks the Rhodes Trust, the Natural Sciences and Engineering Research Council of Canada (NSERC), and Schmidt Science Fellows, in partnership with the Rhodes Trust, for financial support. We are grateful to the UK Materials and Molecular Modelling Hub for computational resources, which is partially funded by EPSRC (EP/P020194/1). Part of this work was performed using resources provided by the Cambridge Service for Data Driven Discovery (CSD3) operated by the University of Cambridge Research Computing Service (www.csd3.cam.ac.uk), provided by Dell EMC and Intel using Tier-2 funding from the Engineering and Physical Sciences Research Council (capital grant EP/P020259/1), and DiRAC funding from the Science and Technology Facilities Council (www.dirac.ac.uk). We are also grateful to Philipp Nickels and Tim Liedl for very helpful discussions.

\end{acknowledgement}

\begin{suppinfo}
\begin{itemize}
  \item full simulation details, including further information about the oxDNA thermostats and potentials;
  \item complete derivations for procedures outlined in the main text;
  \item details of the procedure to ensure adequate sampling over secondary structures;
  \item complete, tabulated simulation results; and
  \item additional figures highlighting simulated results.

\end{itemize}

\end{suppinfo}


\providecommand{\latin}[1]{#1}
\makeatletter
\providecommand{\doi}
  {\begingroup\let\do\@makeother\dospecials
  \catcode`\{=1 \catcode`\}=2 \doi@aux}
\providecommand{\doi@aux}[1]{\endgroup\texttt{#1}}
\makeatother
\providecommand*\mcitethebibliography{\thebibliography}
\csname @ifundefined\endcsname{endmcitethebibliography}
  {\let\endmcitethebibliography\endthebibliography}{}


\setcounter{figure}{0}
 \makeatletter
 \renewcommand{\thefigure}{S\@arabic\c@figure}
 \setcounter{equation}{0}
 \renewcommand{\theequation}{S\@arabic\c@equation}
 \setcounter{table}{0}
 \renewcommand{\thetable}{S\@arabic\c@table}
 \setcounter{section}{0}
 \renewcommand{\thesection}{S\@arabic\c@section}

\section*{Supporting Information}

\section{Additional Simulation Details} 
The 160-nucleotide poly-T strands were simulated at 21\si{\celsius} and [Na+]=10\,\si{\Molar}, and their endpoints were held in 285\,\si{\pnnm} 3D harmonic traps whose centres were 41.9\,\si{\nm} apart. For all simulations, the oxDNA parameters shown in Table ~\ref{tab:parameters} were used.
\begin{table}[]
\caption{oxDNA parameters used across all simulations described in this work, with the exception of those which used a Langevin thermostat (detailed in the main text), for which \texttt{thermostat = langevin}.}
\begin{tabular}{ll}
\hline
\textbf{Parameter} & \textbf{Value} \\ \hline
newtonian\_steps   & 103            \\
diff\_coef         & 2.50           \\
thermostat         & john         \\
dt                 & 0.005          \\
verlet\_skin       & 0.2            \\
use\_edge          & false          \\ \hline
\end{tabular}
\label{tab:parameters}
\end{table}
\subsection{Langevin thermostat}
oxDNA provides the option of using a Langevin thermostat, which mimics coupling to a thermal bath through the addition of a frictional damping term and a random forcing term to Newton's equations of motion~\cite{book:allen2017}. The strength of the random forces is related to the magnitude of the frictional damping through the `fluctuation-dissipation relation', which guarantees evolution in the canonical NVT ensemble~\cite{book:allen2017}. Because oxDNA simulations are somewhat slower when the Langevin, rather than Andersen-like, thermostat is used, it makes only a brief appearance in the main text.
\subsection{Derivation of FENE force distribution}
The expected force probability distribution for a polymer chain with only backbone interactions implemented can be derived by considering the finitely extensible nonlinear elastic (FENE) potential in oxDNA, which is given by

\begin{equation}
V(r) = -\frac{\epsilon}{2}\ln{\Bigg(1-\Big(\frac{r-r_0}{\Delta}\Big)^2\Bigg)},
\label{eq:V_FENE}
\end{equation}

\noindent where $\Delta = 0.25$ (0.21 nm); $\epsilon = 2$ k$_B$T* with T* a reduced temperature, 3000K; and $r_0 = 0.7564 $ (0.644 nm) for oxDNA 2.0~\cite{Snodin2015supp} in its internal unit system. This potential is shown in Figure~\ref{Ni-fig:potentials}. The probability of finding bases separated by $r$ is 

\begin{equation}
P(r) = \frac{e^{-\beta V(r)}}{\int e^{-\beta V(r)} dr} = \frac{e^{-\beta V(r)}}{Z},
\label{eq:PR_FENE}
\end{equation}

\noindent and the free energy, $A$, as a function of $r$ is

\begin{equation}
A(r) = - k_B T \ln(P(r)).
\label{eq:A_FENE}
\end{equation}

\noindent The force experienced by nucleotides separated by distance $r$ is

\begin{equation}
\vec{F}(r) = - \frac{\partial A}{\partial r} \hat{\textbf{r}} = - \frac{\partial V}{\partial r} \hat{\textbf{r}} = \frac{\frac{\epsilon}{\Delta^2} (r-r_0) }{1-\big(\frac{r-r_0}{\Delta}\big)^2} \hat{\textbf{r}},
\label{eq:Fr_FENE}
\end{equation}

\noindent and is directed along the backbone. If we let $\vec{F}(r) = F\hat{\textbf{r}}$, the probability density function for observing a force $F$, $g(F)$, can be found by noting $g(F) \, dF = P(r(F)) \, dr$, or $g(F) = P(r(F)) \, \frac{dr}{dF}$, where $P(r)$ is given by equation \ref{eq:PR_FENE} and $r(F)$ is found by inverting equation \ref{eq:Fr_FENE}:

\begin{equation}
r(F) = -\frac{\epsilon}{2F} \pm \frac{1}{2F}\sqrt{\epsilon^2 + 4\Delta^2F^2} + r_0.
\label{eq:r_f_FENE}
\end{equation}

\noindent Only the solution featuring the negative square root is bounded. Therefore,

\begin{equation}
g(F) = P(r(F)) \frac{dr}{dF} = \frac{1}{Z} \Bigg[ 1 - \frac{1}{\Delta^2}\bigg( -\frac{\epsilon}{2F} - \frac{1}{2F}\sqrt{\epsilon^2 + 4\Delta^2F^2} \bigg)^2 \Bigg]^{\epsilon \beta/2}\frac{dr}{dF}.
\label{eq:g_f_FENE}
\end{equation}

\noindent From equation~\ref{eq:r_f_FENE},

\begin{equation}
\frac{dr}{dF} = \frac{\epsilon}{2F^2} + \frac{2\Delta^2}{\sqrt{\epsilon^2 + 4\Delta^2F^2}} - \frac{\sqrt{\epsilon^2 + 4\Delta^2F^2}}{2F^2}.
\label{eq:rprime_f_FENE}
\end{equation}

\noindent While keeping track of force components $F_x$, $F_y$, $F_z$ (and thus force magnitude, $|F|$) is straightforward in oxDNA, the vector direction $\hat{\textbf{r}}$ is not as easily accessible, and indeed will be different for all nucleotides along the chain. Therefore, for simplicity, instead of working with $g(F)$ (where $F$ is defined by $\vec{F}(r) = F\hat{\textbf{r}}$), we derive $P(F_z)$, the probability distribution for a single component of the force. To do so, we note that for a random variable $A$ defined as the product of two other random variables, $A=BC$, the probability density distributions of each of the three variables are related thus~\cite{book:rohatgi1976}:

\begin{equation}
P_A(a) = \int_{-\infty}^{\infty} P_B(b)P_C(a/b)\frac{1}{|b|} \, db.
\label{eq:productdist}
\end{equation}

\noindent Since $F_z = \vec{F}  \cdot \hat{\textbf{z}} = F\cos{\theta}$, we can use equation~\ref{eq:productdist} with $A=F_z$, $B=F$, and $C=\cos{\theta}$ to obtain

\begin{equation}
P_{F_z}(f_z) = \int_{-\infty}^{\infty} g_F(f')P_{cos\theta}(f_z/f')\frac{1}{|f'|} \, df'
\label{eq:product_revised}
\end{equation}

\noindent We assume for simplicity that the instantaneous force vector is equally likely to point in all directions; namely, we assume $P(\phi,\theta) = \frac{1}{4\pi}$. In this isotropic case, $P(\theta) = \frac{1}{2}\sin{\theta}$, and the probability density distribution for $\cos{\theta}$ is given by

\begin{equation}
P_{cos\theta}(\cos{\theta}) = P(\theta)\Big|\frac{d\theta}{d(\cos{\theta})}\Big| = \frac{1}{2}.
\end{equation}

\noindent Now, we adjust the integration limits of Eq.~\ref{eq:product_revised} to reflect the fact that $F_z \le F$ by definition:

\begin{equation}
P(F_z)= \int_{-\infty}^{-F_z} \frac{g_F(f')}{2 |f'|} \, df' + \int_{F_z}^\infty \frac{g_F(f')}{2 |f'|} \, df'
\label{eq:finalFENE}
\end{equation}
\noindent Eq.~\ref{eq:finalFENE} is plotted in black in Fig.~\ref{fig:ssDNAexample1}(c) in the main text. The agreement with simulation data is excellent where stacking interactions are suppressed, as expected. 
\subsection{Stacking potential}
To discover why the stacking interaction broadens the force distributions in Fig.~\ref{fig:ssDNAexample1}(c) in the main text, we can make a simple argument based on the form of the stacking potential in oxDNA. 

For simplicity, we consider only the radial part of the stacking potential:
\begin{equation}
V(r) = \epsilon_s\Big(1 - e^{-a(r-r_{0s})}\Big)^2 - \epsilon_s\Big(1 - e^{-a(r_c-r_{0s})}\Big)^2,
\label{eq:stack_r}
\end{equation}

\noindent where $\epsilon_s=1.3523 + 2.6717k_BT$, $r_c = 0.9$, $r_{0s} = 0.4$, and $a=6$ for oxDNA 2.0~\cite{Snodin2015supp} in its internal unit system.

\begin{figure}
\begin{center}
\includegraphics[width=0.9\textwidth]{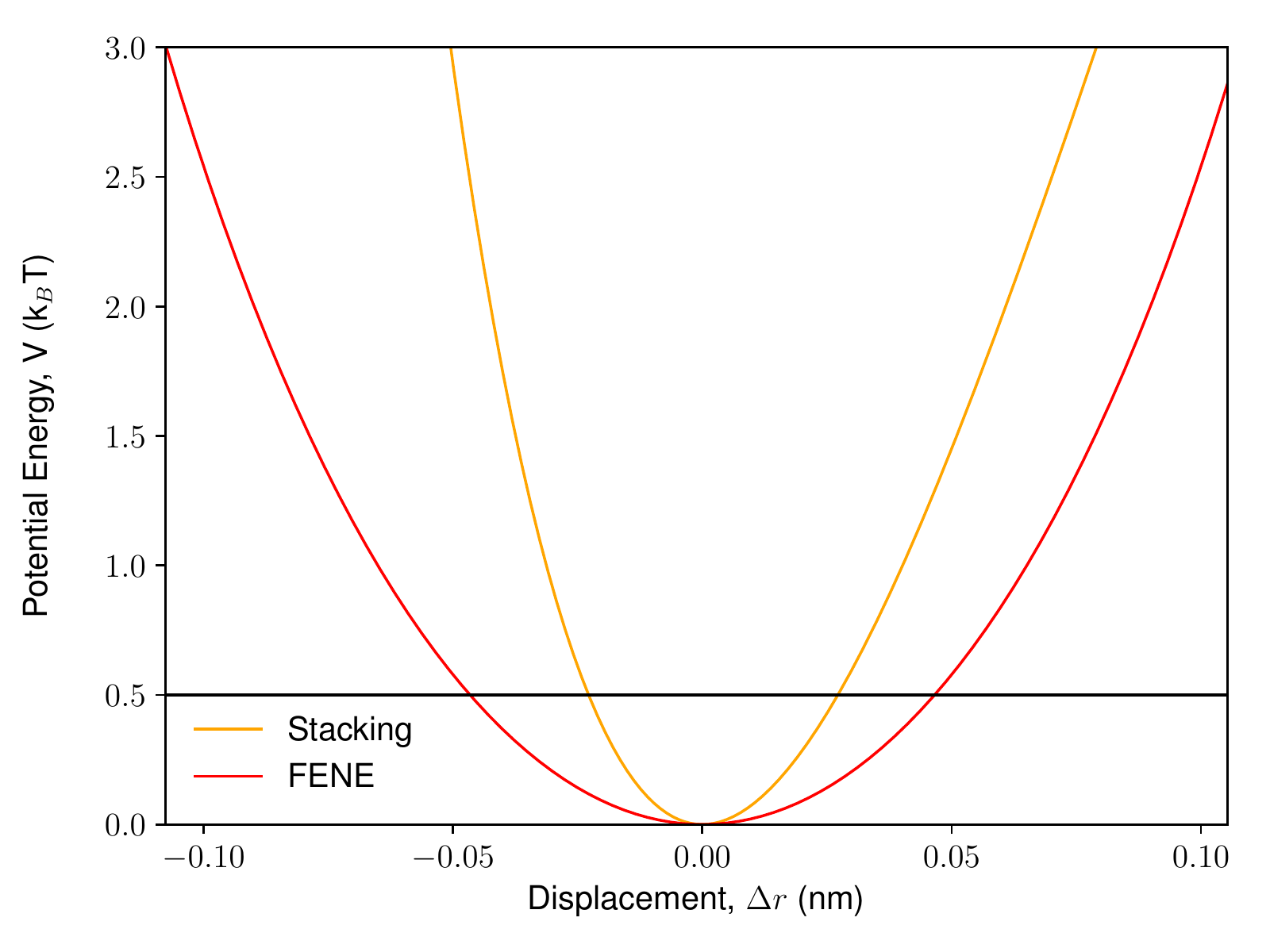}
\caption{oxDNA2.0 radial FENE (red) and stacking (orange) potentials, with the minima of the potentials shifted to the origin. The stacking potential is clearly much steeper than the FENE at the thermal energy (black line), giving rise to the large force distribution width for the stacking curves in Fig.~\ref{fig:ssDNAexample1}(c) of the main text.}
\label{Ni-fig:potentials}
\end{center}
\end{figure}


Figure~\ref{Ni-fig:potentials} shows this potential alongside the radial part of the FENE potential, with the minima of the potentials shifted to the origin. The black line indicates the thermal energy that the system is expected to possess in equilibrium according to equipartition, i.e. $k_BT/2$, at  T=21\,\si{\celsius}. We can evaluate the force, ${F(\Delta r)} = -\frac{dV}{dr}$, at the intersection of each potential with the thermal energy, which occurs at radial displacements from equilibrium of $\Delta r=0.13$\,\si{\nm} and $\Delta r=0.07$\,\si{\nm} for the FENE and stacking potentials, respectively. $F^{thermal}_{FENE} \sim 89$ pN and $F^{thermal}_{stack} \sim 193$ pN for these displacements, which, after taking into account that only one component of the force is plotted in Fig.\,\ref{fig:ssDNAexample1}(c) of the main text, match the observed distribution widths well. We can thus confidently attribute the form of the observed force distributions to the oxDNA potentials, and the substantial force distribution width associated with stacking interactions to the steepness -- and thus, large instantaneous forces -- of the stacking potential.
\subsection{Calculation of effective trap stiffness}
Consider two nucleotides separated along the $z$ axis -- the long axis of the force clamp -- by $d$. We assume each nucleotide moves in an isotropic 3D harmonic potential for simplicity, such that the Boltzmann probability of observing position $r_1$ for nucleotide 1, positioned at the origin, is 

\begin{equation}
\begin{split}
P(\mathbf{r_1}) = P(x_1,y_1,z_1) &= \frac{e^{-\beta k(x_1^2+y_1^2+z_1^2)/2}}{\int_{-\infty}^{\infty}e^{-\beta k(x_1^2+y_1^2+z_1^2)/2}dx_1\,dy_1\,dz_1} \\
&= \left(\frac{\beta k}{2\pi}\right)^{3/2}e^{-\beta k(x_1^2+y_1^2+z_1^2)/2}
\end{split}
\end{equation}
\noindent and for nucleotide 2 is
\begin{equation}
\begin{split}
P(\mathbf{r_2}) = P(x_2,y_2,z_2) = \left(\frac{\beta k}{2\pi}\right)^{3/2}e^{-\beta k(x_2^2+y_2^2+(z_2-d)^2)/2},
\end{split}
\end{equation}
\noindent where $\beta$ is the inverse thermal energy and $k$ is the stiffness of each trap. The distance between the two nucleotides is $r = \sqrt{(x_2-x_1)^2+(y_2-y_1)^2+(z_2-z_1)^2}$, so the mean $r$ and $r^2$ values are given by (where $dV_i = dx_idy_idz_i$):
\begin{equation}
\langle r \rangle = \int_{-\infty}^{\infty} \sqrt{(x_2-x_1)^2+(y_2-y_1)^2+(z_2-z_1)^2}P(x_1,y_1,z_1)P(x_2,y_2,z_2) dV_{1}dV_{2}
\label{eq:rmean}
\end{equation}
\begin{equation}
\langle r^2 \rangle = \int_{-\infty}^{\infty} \big((x_2-x_1)^2+(y_2-y_1)^2+(z_2-z_1)^2\big)P(x_1,y_1,z_1)P(x_2,y_2,z_2)  dV_{1}dV_{2}
\label{eq:r2}
\end{equation}

\noindent where the joint probability distribution $P(x_1,y_1,z_1,x_2,y_2,z_2)$ is simply the product \\
$P(x_1,y_1,z_1)P(x_2,y_2,z_2)$ since the variables are independent. Equations~\ref{eq:rmean} and \ref{eq:r2} give the mean and variance ($\langle r^2 \rangle - \langle r \rangle^2$) of the $P(r)$ distribution, and can be numerically integrated with different values of $k$ until the observed parameters are obtained. Doing so yields an effective trap stiffness of $k=16.32$\,\si{\pnnm} and separation of $d=42.76$\,\si{\nm}. These values are only slightly different from the results obtained by neglecting position fluctuations in $x$ and $y$ altogether, assuming $r=\Delta z=z_2-z_1$, and taking the variance of $P(r)$ to be the sum $\sigma_r^2 = \sigma_{z_1}^2 + \sigma_{z_2}^2 = \frac{1}{\beta k} + \frac{1}{\beta k}$, with the individual variances $\sigma_{z_i}^2$  equal and determined using the equipartition theorem. Under these assumptions, $k=16.23$\,\si{\pnnm} and $d=42.77$\,\si{\nm}. For origami geometries with smaller $z:x$ ratios, however, this more accurate method for calculating effective stiffness may prove necessary.

\section{Additional Results} 
\begin{figure}
\begin{center}
\includegraphics[scale=0.75]{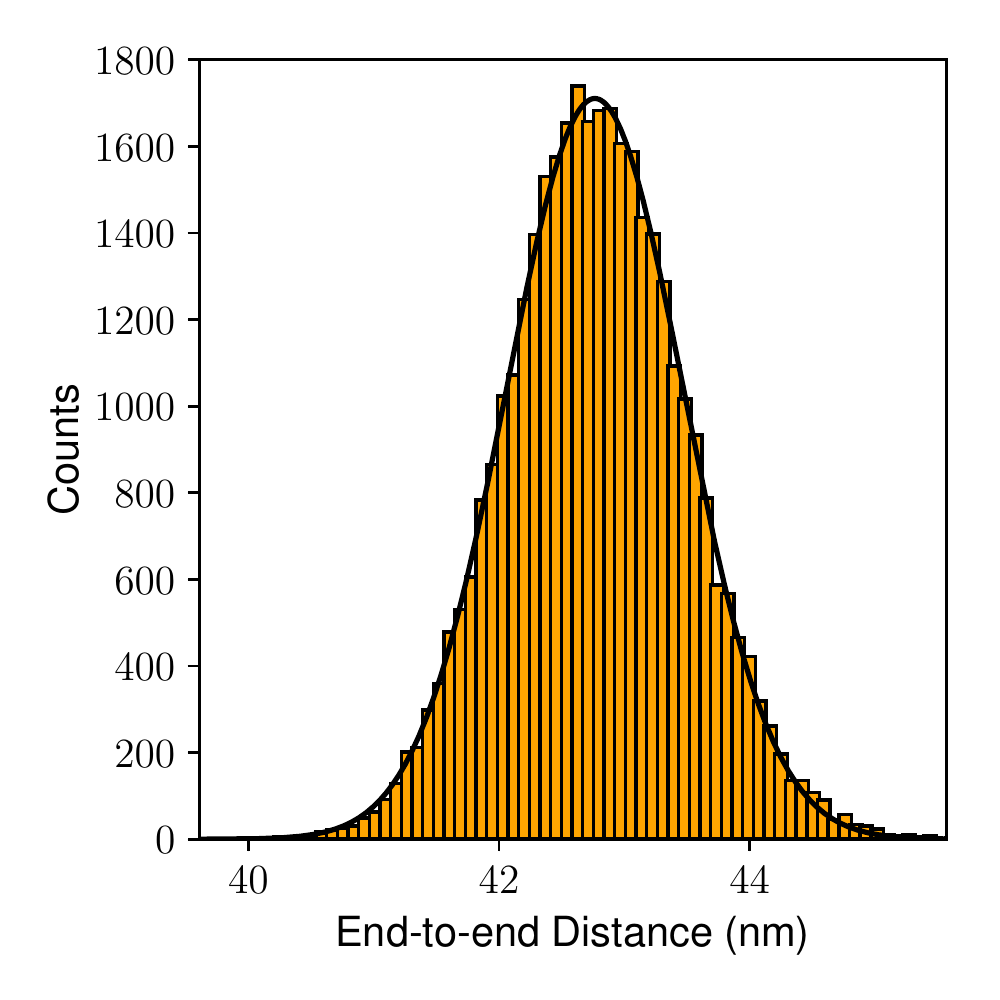}
\caption{End-to-end distribution for scaffold nucleotides on either edge of the central gap in the structure of \citet{Nickels2016supp}. The width of this distribution can be used to determine an effective stiffness for the traps that are used to represent the constraints that the origami places on the tension-bearing single strand, as detailed in the main text.}
\label{fig:gauss_ree}
\end{center}
\end{figure}

\begin{figure}
\begin{center}
\includegraphics[width=0.9\textwidth]{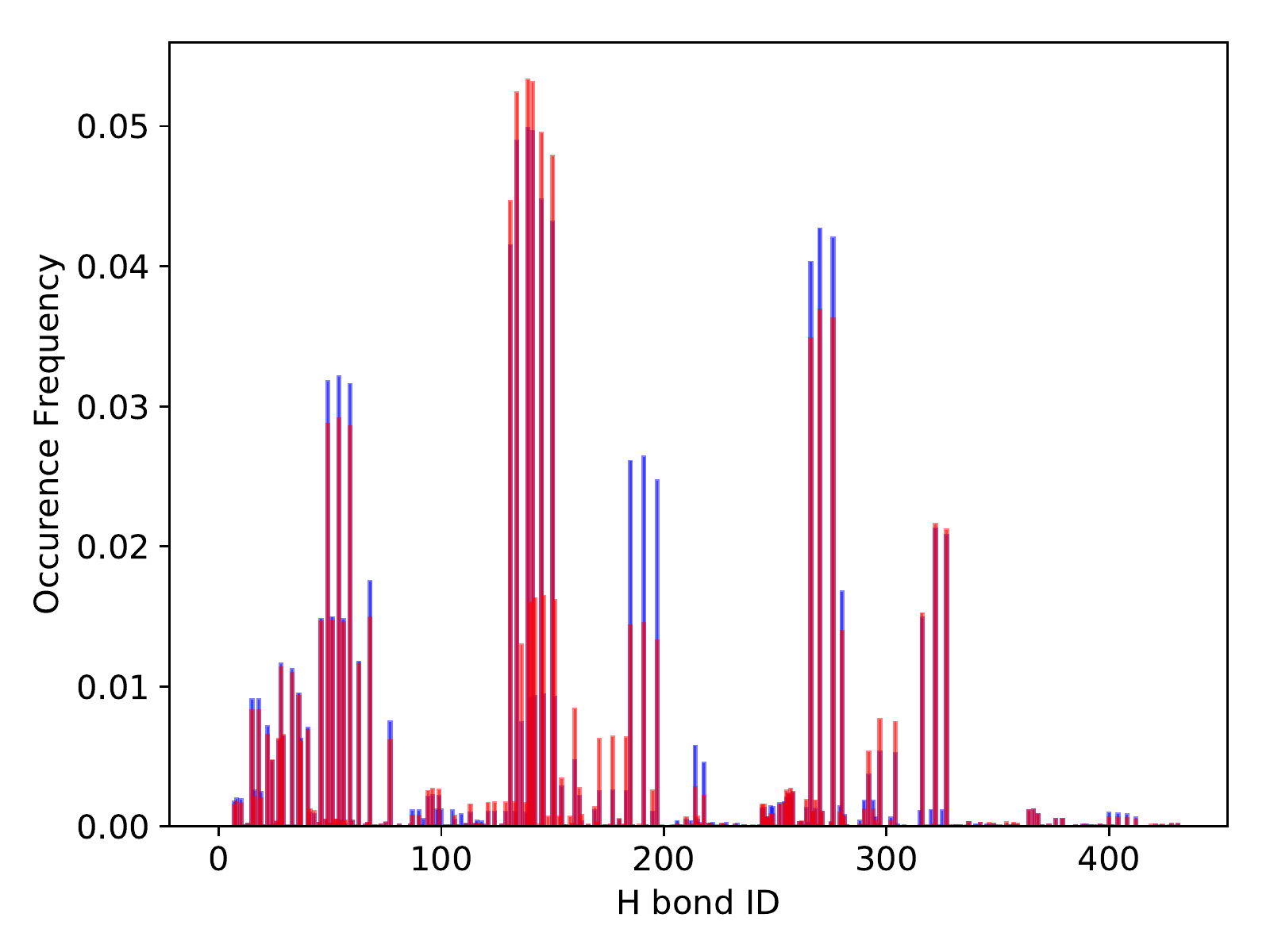}
\caption[]{Map of all unique base pairs observed in oxDNA simulations of the 4.0\,\si{\pn} IsoII structure at [Na$^{+}$]=5\,\si{\Molar}. The data were split in half (12 identical replicas of 1.3$\times10^9$ MD steps each), indicated by red and blue histograms. The amount of overlap gives a rough indication of whether the important secondary structures are being sampled sufficiently.}
\label{Ni-fig:4sec_struct}
\end{center}
\end{figure}
\begin{figure}
\begin{center}
\includegraphics[width=0.9\textwidth]{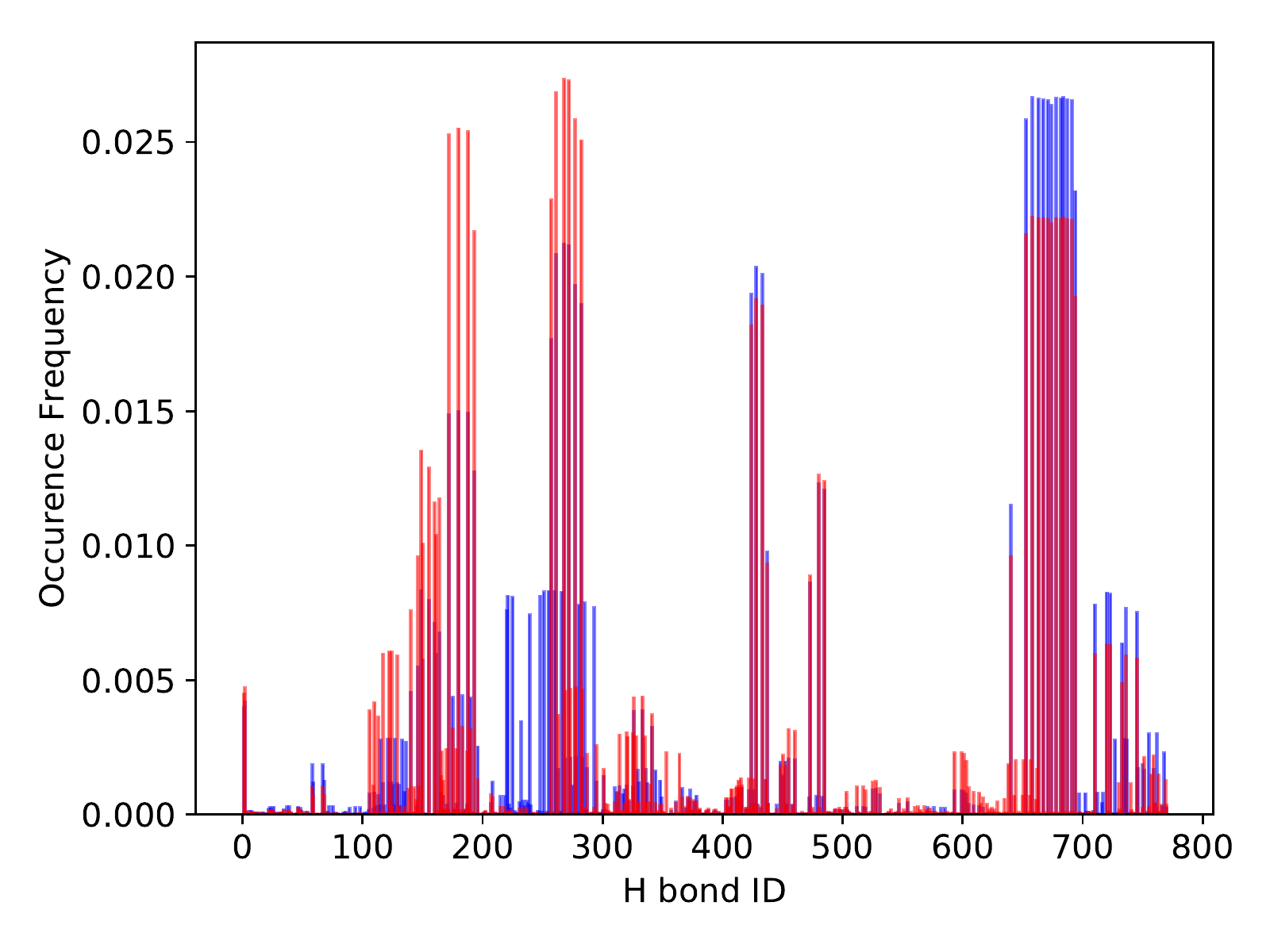}
\caption[]{Map of all unique base pairs observed in oxDNA simulations of the 2.5\,\si{\pn} IsoII structure at [Na$^{+}$]=5\,\si{\Molar}. The data were split in half (12 identical replicas of 1.4$\times10^9$ MD steps each), indicated by red and blue histograms. The amount of overlap gives a rough indication of whether the important secondary structures are being sampled sufficiently.}
\label{Ni-fig:2-5sec_struct}
\end{center}
\end{figure}
\begin{figure}
\begin{center}
\includegraphics[width=0.9\textwidth]{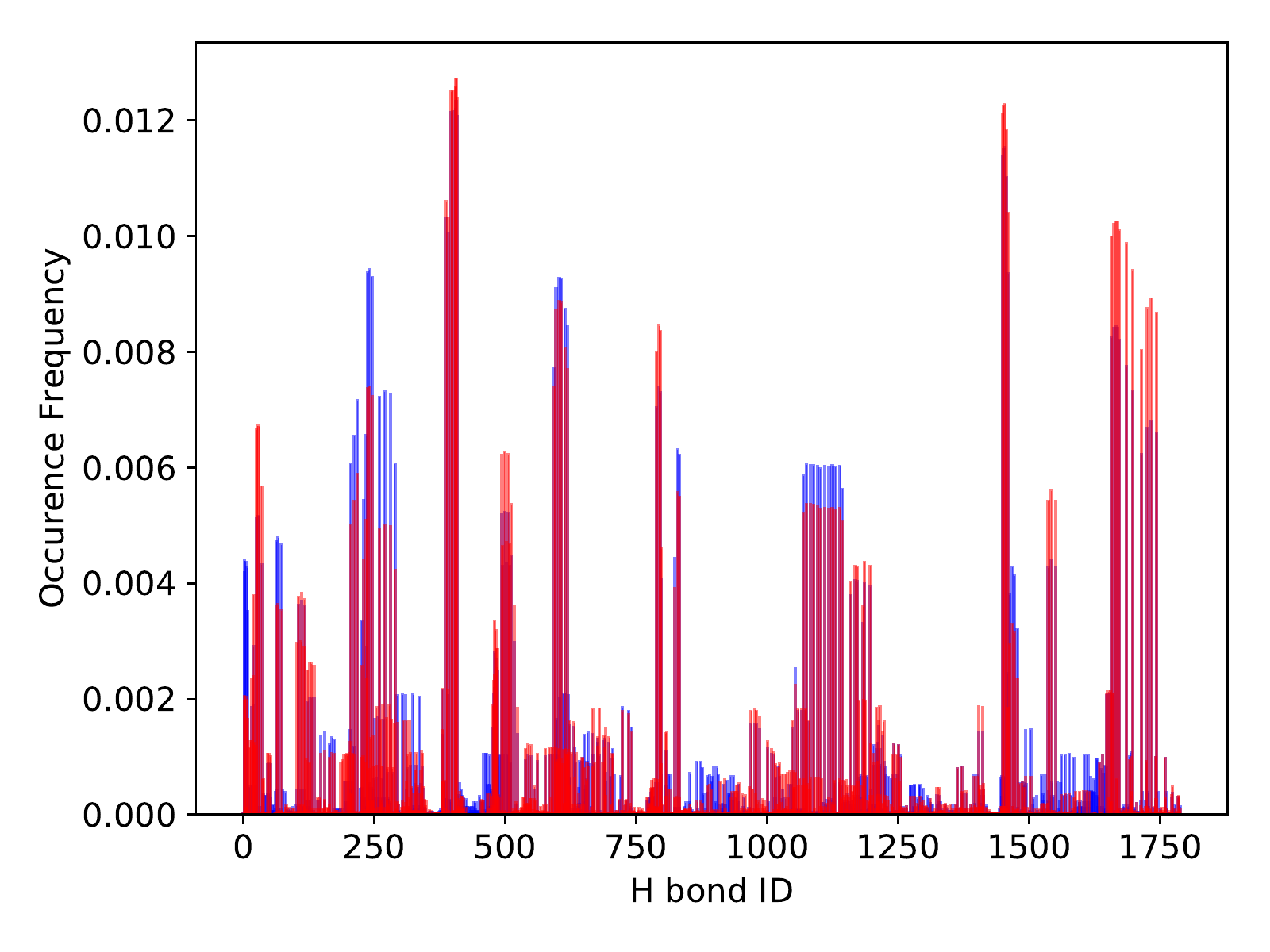}
\caption[]{Map of all unique base pairs observed in oxDNA simulations of the 1.2\,\si{\pn} IsoII structure at [Na$^{+}$]=5\,\si{\Molar}. The data were split in half (12 identical replicas of 8$\times10^8$ MD steps each), indicated by red and blue histograms. The amount of overlap gives a rough indication of whether the important secondary structures are being sampled sufficiently.}
\label{Ni-fig:1-2sec_struct}
\end{center}
\end{figure}

As explained in the main text, it is important to ensure that our simulations are sampling all relevant secondary structures. To gauge whether we had sampled secondary structures sufficiently, we split our simulation output for each system in half and computed histograms of which bases formed hydrogen bonds throughout the simulation, shown in Figures~\ref{Ni-fig:4sec_struct}-\ref{Ni-fig:1-2sec_struct} for all three designed forces. Only IsoII results at the highest salt concentration simulated, 5\,\si{\Molar} [Na$^{+}$], are pictured, as secondary structures become less favourable at lower ionic strengths due to electrostatic repulsion and secondary structure sampling is therefore most difficult at high salt. Though well over 10\,000 possible pairs of bases exist for these strands, the figures show only those hydrogen bonds which were actually observed and assigns a numeric ID to each unique bond. The amount of overlap between blue and red histograms is taken as a proxy of how well secondary structures were sampled. For the 4.0\,\si{\pn} design, the main patterns of secondary structures visited are the same for both histograms, and we conclude that $\sim$1$\times10^9$ MD steps ($\times$ 24 replicas) is an appropriate simulation length for this system. We expect equilibration over secondary structures to be easier at higher internal forces, since force helps to destabilize secondary structures. Indeed, the histogram overlap is less for the larger 2.5\,\si{\pn} and 1.2\,\si{\pn} designs, and more different pairs of bases are seen to interact via hydrogen bonding as the designed force decreases (there are also combinatorially more possibilities for the longer strands). The results reported for these low-force systems should be treated correspondingly more cautiously.  While the accuracy of the results can be improved with longer simulations, we observed the same qualitative trends for all three force designs, and thus have confidence in the results even for the 2.5\,\si{\pn} and 1.2\,\si{\pn} systems.

\begin{figure}
\begin{center}
\includegraphics[width=0.85\textwidth]{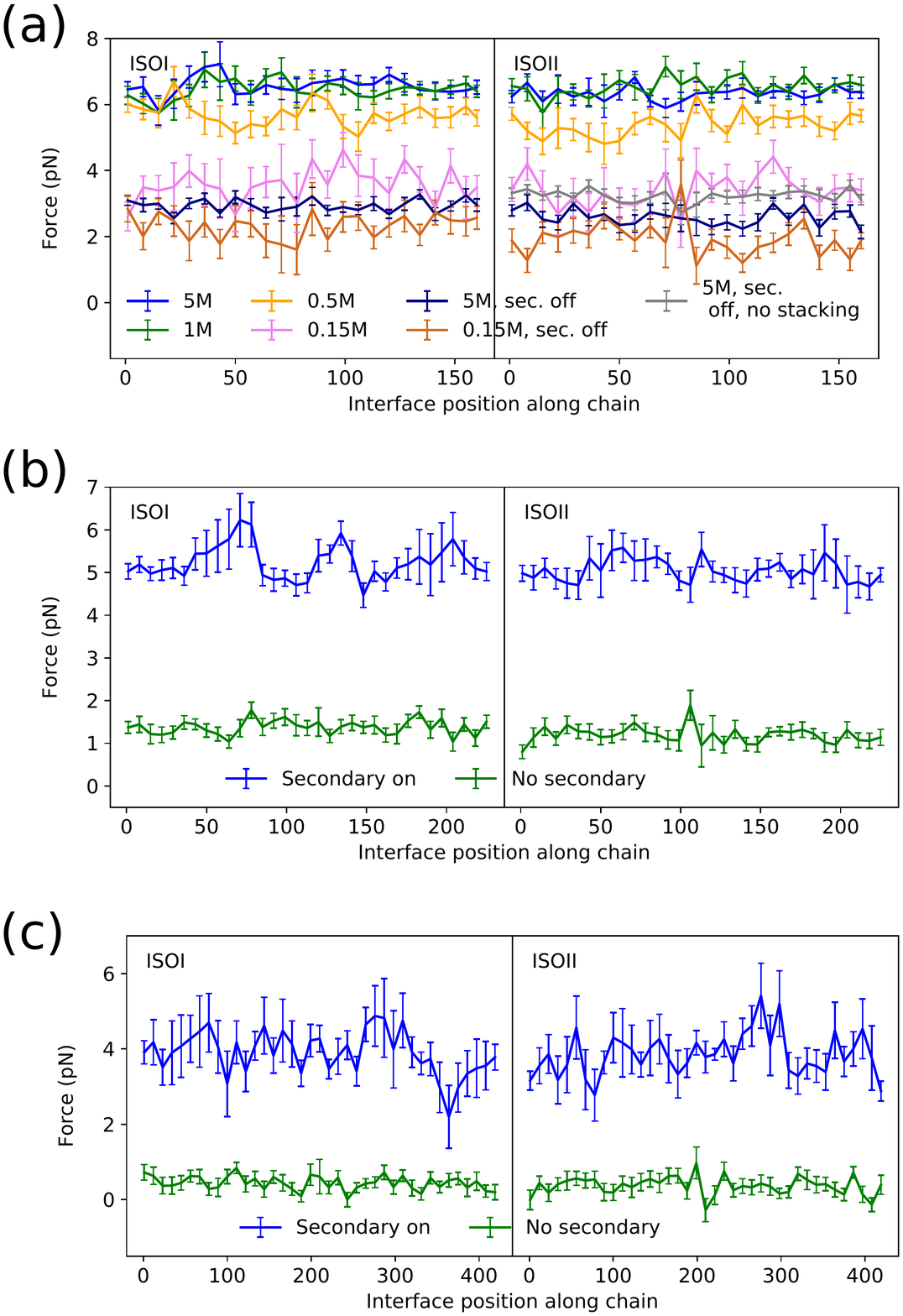}
\caption[]{Net forces crossing interfaces along the ssDNA strand for \citet{Nickels2016supp}'s (a) 4.0\,\si{\pn}; (b) 2.5\,\si{\pn}; and (c) 1.2\,\si{\pn} designs. Errors are standard deviations over $\sim24$ identical replica simulations of $\sim10^9$ MD steps. Results for the 4.0\,\si{\pn} system are shown under various salt conditions, and additionally for simulations in which both secondary structure and stacking interactions are suppressed (grey curve); the forces in this case are higher than when stacking interactions are permitted (indigo curve) because a stacked strand has a longer Kuhn length than an unstacked one, and thus has fewer conformations available to it, lowering the internal force.}
\label{Ni-fig:force_v_interface}
\end{center}
\end{figure}

\begin{figure}
\begin{center}
\includegraphics[width=0.9\textwidth]{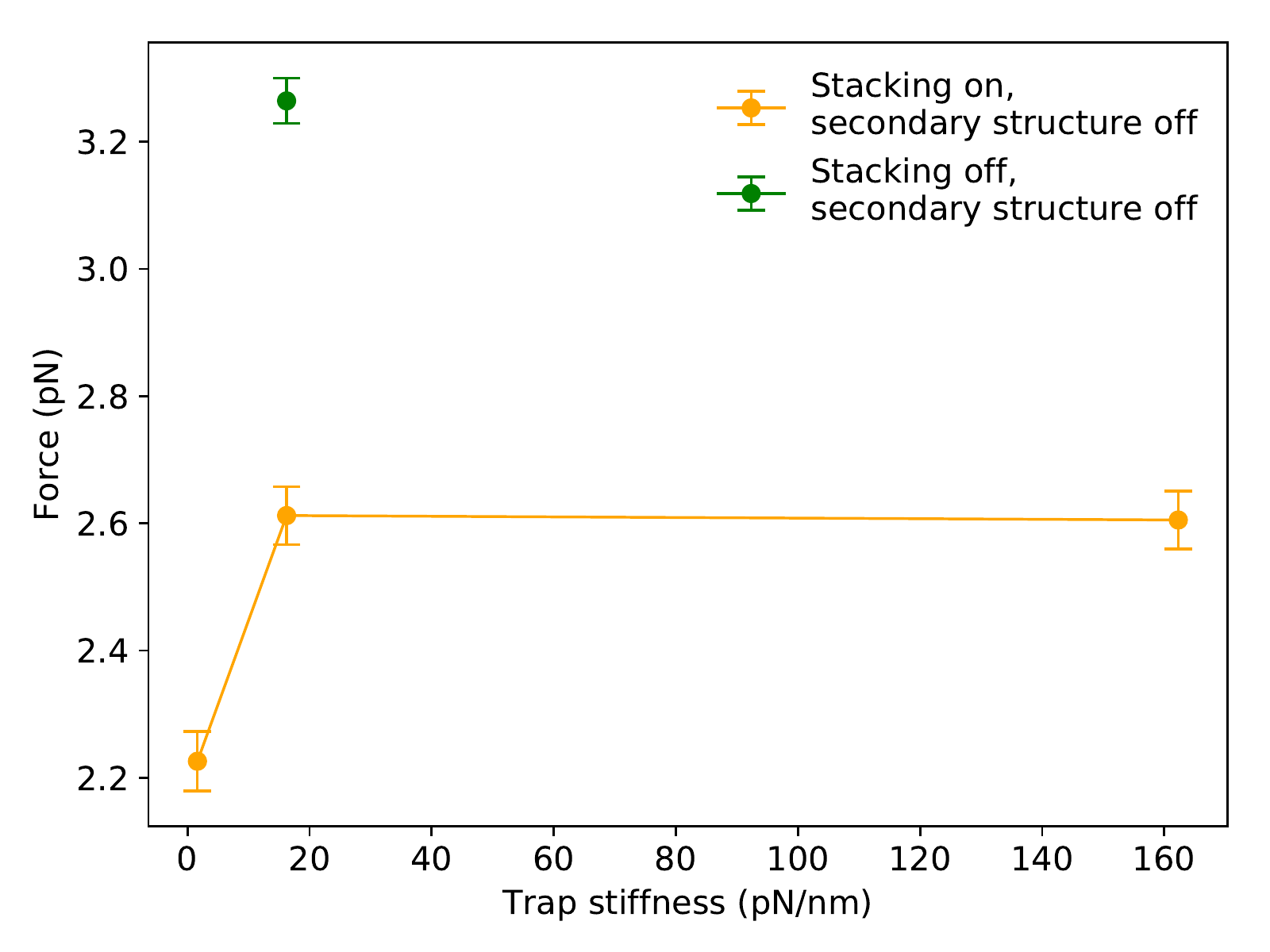}
\caption[]{Effect of trap stiffness on internal force in ssDNA, for the 4.0\,\si{\pn} design of \citet{Nickels2016supp}. Our simulations are performed in the regime where the force has plateaued, and origami stiffness is therefore not expected to be relevant.}
\label{Ni-fig:f_v_stiff}
\end{center}
\end{figure}

The complete results for all systems simulated can be found in Table~\ref{tab:results}. Figure~\ref{Ni-fig:force_v_interface} contains the forces measured at various interfaces along the strand containing the Holliday junction at different salt concentrations, for all three force designs, and with and without secondary structures permitted. The results of simulations of the 4.0pN design with different harmonic trap stiffnesses are summarised in Figure~\ref{Ni-fig:f_v_stiff}; above $\sim20\si{\pn\nm}$, the stiffness doesn't appreciably affect the force, so the assumption of fixed endpoints is reasonable. However, for other origami structures, the compliance may be relevant. 

\begin{table}[]
\caption{Summary of the simulations performed on the system of \citet{Nickels2016supp}, including the average ssDNA internal force in each case. Errors in force are the standard errors of a weighted mean across a number of different nucleotides at which forces were measured, as shown in Fig.~\ref{Ni-fig:force_v_interface}. For almost all sets of parameters, 24 identical replica simulations were run; where fewer replicas are reported below, some runs were discarded due to the (rare) occurrence of a transition between isomer states.}
\begin{tabular}{|l|l|l|l|l|l|l|l|}
\hline
\textbf{Design} & \textbf{Isomer} & \textbf{\begin{tabular}[c]{@{}l@{}}Secondary\\ structures\end{tabular}} & \textbf{\begin{tabular}[c]{@{}l@{}}k\\ (pN/nm)\end{tabular}} & \textbf{\begin{tabular}[c]{@{}l@{}}Salt\\ (M)\end{tabular}} & \textbf{\begin{tabular}[c]{@{}l@{}}\# MD\\ steps\\ (x24)\end{tabular}} & \textbf{\begin{tabular}[c]{@{}l@{}}approx.\\ CPU\\ days\end{tabular}} & \textbf{\begin{tabular}[c]{@{}l@{}}Force\\ (pN)\end{tabular}} \\ \hline
4.0\,\si{\pn}         & IsoI            & On                                                                      & 16.32                                                        & 5                                                           & \begin{tabular}[c]{@{}l@{}}$1\times10^9$\\ ($\times$23)\end{tabular}   & 85                                                                    & 6.54$\pm$0.05                                                 \\ \cline{5-8} 
                &                 &                                                                         &                                                              & 1                                                           & $9\times10^8$                                                          & 113                                                                   & 6.46$\pm$0.06                                                 \\ \cline{5-8} 
                &                 &                                                                         &                                                              & 0.5                                                         & $5.8\times10^8$                                                        & 96                                                                    & 5.71$\pm$0.07                                                 \\ \cline{5-8} 
                &                 &                                                                         &                                                              & 0.15                                                        & \begin{tabular}[c]{@{}l@{}}$4.8\times10^8$\\ ($\times$19)\end{tabular} & 102                                                                   & 3.6$\pm$0.1                                                   \\ \cline{3-8} 
                &                 & Off                                                                     & 16.32                                                        & 5                                                           & $1.3\times10^9$                                                        & 104                                                                   & 2.91$\pm$0.03                                                 \\ \cline{5-8} 
                &                 &                                                                         &                                                              & 0.15                                                        & \begin{tabular}[c]{@{}l@{}}$5\times10^8$\\ ($\times$19)\end{tabular} & 103                                                                  & 2.33$\pm$0.09                                                           \\ \cline{2-8} 
                & IsoII           & On                                                                      & 16.32                                                        & 5                                                           & $1.3\times10^9$                                                        & 113                                                                   & 6.35$\pm$0.04                                                 \\ \cline{5-8} 
                &                 &                                                                         &                                                              & 1                                                           & $9\times10^8$                                                          & 113                                                                   & 6.52$\pm$0.05                                                 \\ \cline{5-8} 
                &                 &                                                                         &                                                              & 0.5                                                         & $5.8\times10^8$                                                        & 96                                                                    & 5.43$\pm$0.06                                                 \\ \cline{5-8} 
                &                 &                                                                         &                                                              & 0.15                                                        & $4.8\times10^8$                                                        & 130                                                                   & 3.4$\pm$0.1                                                   \\ \cline{3-8} 
                &                 & Off                                                                     & 1.632                                                        & 5                                                           & $7.8\times10^8$                                                        & 64                                                                    & 2.23$\pm$0.05                                                 \\ \cline{4-8} 
                &                 &                                                                         & 16.32                                                        & 5                                                           & $7.8\times10^8$                                                        & 64                                                                    & 2.61$\pm$0.05                                                 \\ \cline{5-8} 
                &                 &                                                                         &                                                              & 0.15                                                        & $5.7\times10^8$                                                        & 131                                                                  & 1.88$\pm$0.07                                                        \\ \cline{4-8} 
                &                 &                                                                         & 163.2                                                        & 5                                                           & $8.8\times10^8$                                                        & 64                                                                    & 2.61$\pm$0.05                                                 \\ \cline{3-8} 
                &                 & Off, no stacking                                                        & 16.32                                                        & 5                                                           & $7.8\times10^8$                                                        & 59                                                                    & 3.26$\pm$0.04                                                 \\ \hline
2.5\,\si{\pn}         & IsoI            & On                                                                      & 16.32                                                        & 5                                                           & $1.2\times10^9$                                                        & 148                                                                   & 5.07$\pm$0.05                                                          \\ \cline{3-8} 
                &                 & Off                                                                     & 16.32                                                        & 5                                                           & \begin{tabular}[c]{@{}l@{}}$1.4\times10^9$\\ ($\times$23)\end{tabular} & 131                                                                   & 1.38$\pm$0.03                                                 \\ \cline{2-8} 
                & IsoII           & On                                                                      & 16.32                                                        & 5                                                           & $1.4\times10^9$                                                        & 169                                                                   & 5.01$\pm$0.05                                                          \\ \cline{3-8} 
                &                 & Off                                                                     & 16.32                                                        & 5                                                           & \begin{tabular}[c]{@{}l@{}}$1.4\times10^9$\\ ($\times$23)\end{tabular} & 131                                                                   & 1.19$\pm$0.03                                                 \\ \hline
1.2\,\si{\pn}          & IsoI            & On                                                                      & 16.32                                                        & 5                                                           & $8\times10^8$                                                          & 180                                                                   & 3.83$\pm$0.08                                                           \\ \cline{3-8} 
                &                 & Off                                                                     & 16.32                                                        & 5                                                           & \begin{tabular}[c]{@{}l@{}}$1\times10^9$\\ ($\times$23)\end{tabular}   & 155                                                                   & 0.43$\pm$0.03                                                 \\ \cline{2-8} 
                & IsoII           & On                                                                      & 16.32                                                        & 5                                                           & $8\times10^8$                                                          & 180                                                                   & 3.71$\pm$0.07                                                           \\ \cline{3-8} 
                &                 & Off                                                                     & 16.32                                                        & 5                                                           & $7.5\times10^8$                                                        & 40                                                                    & 0.36$\pm$0.03                                                 \\ \hline
\end{tabular}
\label{tab:results}
\end{table}

\providecommand{\latin}[1]{#1}
\makeatletter
\providecommand{\doi}
  {\begingroup\let\do\@makeother\dospecials
  \catcode`\{=1 \catcode`\}=2 \doi@aux}
\providecommand{\doi@aux}[1]{\endgroup\texttt{#1}}
\makeatother
\providecommand*\mcitethebibliography{\thebibliography}
\csname @ifundefined\endcsname{endmcitethebibliography}
  {\let\endmcitethebibliography\endthebibliography}{}

\end{document}